\documentclass[twocolumn,showpacs,preprintnumbers,amsmath,amssymb,pra,aps,superscriptaddress,nofootinbib]{revtex4-1}

\usepackage{array}
\usepackage{graphicx}
\usepackage{subfigure}
\usepackage{epsfig}
\usepackage{verbatim}
\usepackage{color}
\usepackage{multirow}

\newcommand{\minitab}[2][l]{\begin{tabular}{#1}#2\end{tabular}}

\begin{document}

\title{Integrated Analysis of Performance and Resource of\\ Large-Scale Quantum Computing}
\author{Yongsoo Hwang}
\author{Taewan Kim}
\author{Chungheon Baek}
\author{Byung-Soo Choi}

\affiliation{Electronics and Telecommunications Research Institute}

\begin{abstract}
To see the feasibility of a large-scale quantum computing, it is required to accurately analyze the performance and the quantum resource.
However, most of the analysis reported so far have focused on the statistical examination, i.e., simply calculating the performance and resource based on individual data, and even worse usually only a few components have been considered.
In this work, to achieve more exact analysis, we propose an integrated analysis method for a practical quantum computing model with three components (\textit{algorithm}, \textit{error correction} and \textit{device}) under a realistic quantum computer system architecture.
To implement the above method, we develop a quantum computing framework composed of three functional layers: compile, system and building block. 
This framework can support, for the first time, the mapping of quantum algorithm from physical qubit level to system  architecture level with a given fault-tolerant scheme.
Therefore, the proposed method can measure the effect of dynamic situation when the quantum computer practically runs.
By using our method, we found that Shor algorithm to factorize 512-bit integer requires $8.78\times 10^ 5$ hours.
We also show how the proposed method can be used for analyzing optimal concatenation level and code distance of fault-tolerant quantum computing.
\end{abstract}


\maketitle

\section{Introduction}\label{sec:introduction}
Over the last decades, diverse quantum computing components from application to physical device have been actively researched and developed.
Some of them already have theoretically optimal performance~\cite{Jones:2013ii,Muralidharan:2016ky,Ross:2016uz,Bombin:2015hi,Napp:2012vn,Bombin:2007ed}.
Gigantic IT corporations have announced that they succeeded in the development of dozens of qubit system~\cite{IBM:Iidq8v80,Kelly:2018uc,Intel:1vYziePb}.
Even they expect to see a thousand qubit system within ten years.
Besides, several startups also started to develop a quantum computing hardware and software~\cite{QuantumComputingReport:FR0ihxsu}.
It seems that the era of a quantum computing is gradually coming close to us.

So far, to see the feasibility of a practical large-scale quantum computing, some efforts have been dedicated to investigation into the quantity of the required quantum resource~\cite{Suchara:2013ho,Smith:2014ux,Reiher:2017cv,Goto:2016jo,Suchara:2013tg,JavadiAbhari:2015jf,Steiger:2018to} and the expected performance~\cite{Fowler:2012fi,Jones:2012kc,VanMeter:2009ut,Suchara:2013tg}.
Based on such analysis, the security of conventional cryptography against a theoretically super-fast quantum computing also has been analyzed~\cite{Grassl:2015wa}.
However, we think that the analysis reported so far are not fully satisfactory because they usually have considered only a few components of a quantum computing and/or focused on statistical examination. 
In particular, the analysis results with the statistical examination are nothing more than simple calculation based on individual data of several quantum computing components~\cite{Suchara:2013ho,Suchara:2013tg}.

To achieve more exact analysis, we believe the practical operation of a quantum computing needs to be considered.
For that, one first has to consider quantum computing components in terms of the practical operation.
For example, we have to prepare a quantum algorithm decomposed into individual gates instead of only the dominant part of a quantum algorithm.
Second, the above components have to be properly integrated with a target quantum computer system architecture.
For example, we have to find a path for qubit movements (or braiding) on the target system architecture without raising any conflicts with other paths.
By doing so, the quantity of qubit movements which cannot be considered in the statistical analysis can be exactly measured, and therefore we can say that the analysis results are realistic.

In this work, to perform the integrated analysis more efficiently, we propose and develop a quantum computing framework composed of three functional layers (compile, system and building block), where each layer has a well-defined role for a quantum computing as follows.
\begin{itemize}
\item Compile: Decomposition of a quantum algorithm into a sequence of quantum gates called a \textit{quantum assembly code}
\item System: Integration of a quantum algorithm and building blocks under a quantum computer system architecture
\item Building Block: Implementation of logical qubits and gates according to a quantum error-correcting code and a fault-tolerant scheme
\end{itemize}
With the proposed framework, we can conduct the analysis in the most fine-grained manner.
For example, we even count a qubit movement one by one, and the quantity of physical qubits based on the provided quantum computer architecture and the ancilla qubits required for error correction and magic state.
Furthermore, the framework examines the performance and the resource with the criterion, 100\% fidelity quantum computing~\cite{Whitney:2009wh}.
By doing so, we can provide full quantum resource and compare various quantum computing configurations fairly.

The objective of this work is to estimate the most accurate quantum computing performance and quantum resource by the help of the framework we developed.
For that, we first need to configure a quantum computing by choosing specific protocols and device properties. 
By doing so, as mentioned above, we can see the quantum resource and the performance of a quantum computing based on most of quantum computing components from algorithm to hardware.
Besides, it is also possible to see which component affects a quantum computing most seriously by applying realistic technology from the top layer to the bottom layer gradually.
In this regards, we can see that a theoretically optimal protocol really leads to optimal quantum computing or instead works suboptimally in composition with other components.

By exploiting the accurate analysis function, we can show numerical data for some theoretical conjectures in the fault-tolerant quantum computing.
In Steane code based quantum computing, increasing a concatenation level makes it possible to achieve more reliable quantum computing, but we guess that the more does not lead to the better all the time because a higher concatenation level requires more quantum resource and longer execution time.
Therefore, it is reasonable to guess that there may exist a trade-off between the reliability and the performance (resource).
In this work, we show there exists such trade-off, and in surface code based quantum computing too.

The remainder of this paper is organized as follows.
Section~\ref{sec:related_works} reviews some related works.
Section~\ref{sec:configuration} overviews the proposed quantum computing framework and describes how to configure and analyze a quantum computing with the proposed framework.
Section~\ref{sec:proposed_framework} describes each layer of the proposed framework in detail.
Section~\ref{sec:performance_metric} describes the performance metric we evaluated in this work, and the analysis results will be shown in section~\ref{sec:performance_evaluation}.
Section~\ref{sec:performance_improvement} discusses, by exploiting the proposed framework, what happens in a quantum computing if an individual component is improved.
This paper concludes with discussions in section~\ref{sec:discussion}.

\section{Related Works}\label{sec:related_works}

\begin{table*}[t]
\caption{
Summary of the related works.
Note that ``$\bigtriangleup$" indicates the corresponding component is partially applied.
Namely, as mentioned before, Refs.~\cite{Fowler:2012fi} and~\cite{Jones:2012kc} perform the performance analysis based on only the dominant part of a quantum algorithm, not covering all quantum gates.
Besides, surface code basically requires that physical qubits are arranged on the 2-dimensional nearest neighbor layout.
Therefore, the related works applying surface code error correction implicitly consider a 2D layout of physical qubits in spite of explicitly no mentioning about the qubit layout.
\newline
}
\small
\centering
\begin{tabular}{c|c|c|c|c|c} \hline 
& \multirow{2}{*}{Compile} & \multirow{2}{*}{FTQC} & Micro-Architecture & System & Analysis \\ 
&  &  & (Qubit Layout) & Synthesis & Criterion\\ \hline
Quipper~\cite{Smith:2014ux,Green:2013ha,Green:2013wb} & \multirow{2}{*}{$\bigcirc$} & \multirow{2}{*}{X} & \multirow{2}{*}{X} & \multirow{2}{*}{X} & {One Time}\\ 
ScaffCC~\cite{JavadiAbhari:2015jf,JavadiAbhari:2014eu,ScaffCCScaffCC:vm} & & & & & Computing \\ \hline
Fowler \textit{et al.}~\cite{Fowler:2012fi} & \multirow{2}{*}{$\bigtriangleup$} & \multirow{2}{*}{Surface} & \multirow{2}{*}{$\bigtriangleup$} & \multirow{2}{*}{X} & {One Time}\\ 
Jones \textit{et al.}~\cite{Jones:2012kc} &  &  &  & & Computing\\ \hline
\multirow{2}{*}{Reiher \textit{et al.}~\cite{Reiher:2017cv}} & \multirow{2}{*}{$\bigcirc$} & \multirow{2}{*}{Surface} & \multirow{2}{*}{$\bigtriangleup$} & \multirow{2}{*}{X} & {One Time}\\ 
 &  &  &  & & Computing \\ \hline
\multirow{3}{*}{QuRE~\cite{Suchara:2013ho,Suchara:2013tg}} & \multirow{3}{*}{$\bigcirc$} & Steane, & \multirow{3}{*}{Layout of Physical Qubits} &\multirow{3}{*}{X} &\multirow{3}{*}{\minitab[c]{One Time\\Computing}}\\ 
 &  & Bacon-Shor, &  & \\ 
 &  & Surface &  & \\ \hline 
\multirow{2}{*}{Present Work} & \multirow{2}{*}{$\bigcirc$} & Steane, & Layouts of Physical and Logical Qubits, & \multirow{2}{*}{$\bigcirc$} & \multirow{2}{*}{Fidelity 100\%}\\ 
 &  & Surface & Communication Bus, Computing Regions & \\ \hline
\end{tabular}
\label{tab:summary_relatedworks}
\end{table*} 

%
We review several quantum computing frameworks discussing the performance and the quantum resource of a quantum computing.
Table~\ref{tab:summary_relatedworks} briefly summarizes the related works.

Quipper~\cite{Smith:2014ux,Green:2013ha,Green:2013wb,Green:2013gi} and ScaffCC~\cite{JavadiAbhari:2015jf,JavadiAbhari:2014eu,ScaffCCScaffCC:vm} are frameworks for quantum compile and resource estimation of a quantum algorithm.
They basically compile a programmed quantum algorithm into a sequence of quantum instructions of a quantum gate and target qubit(s).
From the compile results, they statistically analyze the quantum resource such as the quantities of gates and qubits.
All the data come from a quantum algorithm only.
We believe, without considering physical implementation of the algorithm, the analysis results cannot be used for the reference to the feasibility of a quantum computing.


Fowler \textit{et al.}~\cite{Fowler:2012fi} approximated the quantum computer size and the execution time of Shor algorithm ($N=2000$) with a surface code quantum computing.
However, their analysis is based on the dominant part of the algorithm such as the modular exponentiation circuit only.
Therefore, the execution time completely depends on the quantity of \textit{Toffoli} gates in the circuit, the decomposition of Toffoli gates and measurement gate execution time.
The size of a quantum computer was also calculated based on the quantity of magic states to implement logical \textit{T} gates and the quantity of logical (algorithm) qubits.
Jones~\textit{et al.}~\cite{Jones:2012kc} also estimated the performance and resource of a quantum algorithm within their layered architecture for a quantum computing.
But, their analysis methods are almost the same with Fowler \textit{et al.}~\cite{Fowler:2012fi}.

On the other hand, Reiher \textit{et al.}~\cite{Reiher:2017cv} applied a quantum compile and surface code error correction to estimate the performance and the resource for quantum simulation of complex chemical system.
Since the authors did not consider any quantum computer system architecture and the system integration on the architecture, their analysis corresponds to the statistical calculation based on the data from compiled algorithm and fault-tolerant scheme.
Note that in the literature, the authors claimed that despite overhead of quantum error correction and compile, quantum computations can be performed within reasonable time.

The toolbox for estimating quantum resource \textit{QuRE}~\cite{Suchara:2013ho,Suchara:2013tg} considers most of quantum computing components such as quantum compile, quantum error correction, physical qubit layout and quantum computing device.
By taking quantum compile, they prepared all quantum gates of a quantum algorithm.
Furthermore, they covered both of block-type quantum codes (Steane code and Bacon-Shor code) and surface code and employed 2-dimensional qubit layout for their logical qubits.
In this regards, this toolbox performs the analysis based on more quantum computing components than before.
However, their analysis method is still statistical investigation.
For example, their estimation on the execution time of a quantum computing simply depends on the quantity of quantum gates and its running time, $\sum 1/g_{parallel} \times \#g \times g_T$.
Note that $g_{parallel}$ denotes the quantity of gates $g$ executed in maximally parallel and $g_T$ is the execution time of the gate.
As mentioned before, it is difficult to say their analysis results coincide with the practical situation.


\section{Configuration of Quantum Computing}\label{sec:configuration}
We describe how to configure a quantum computing with the proposed quantum computing framework.
We first overview the structure and functionality of our quantum computing framework, and second describe how to configure a quantum computing.
We also describe which schemes and protocols are currently supported by the framework.

\subsection{Overview of Framework}\label{subsec:overview_framework}
\begin{figure}[t]
\centering
\subfigure[]{
	\epsfig{file=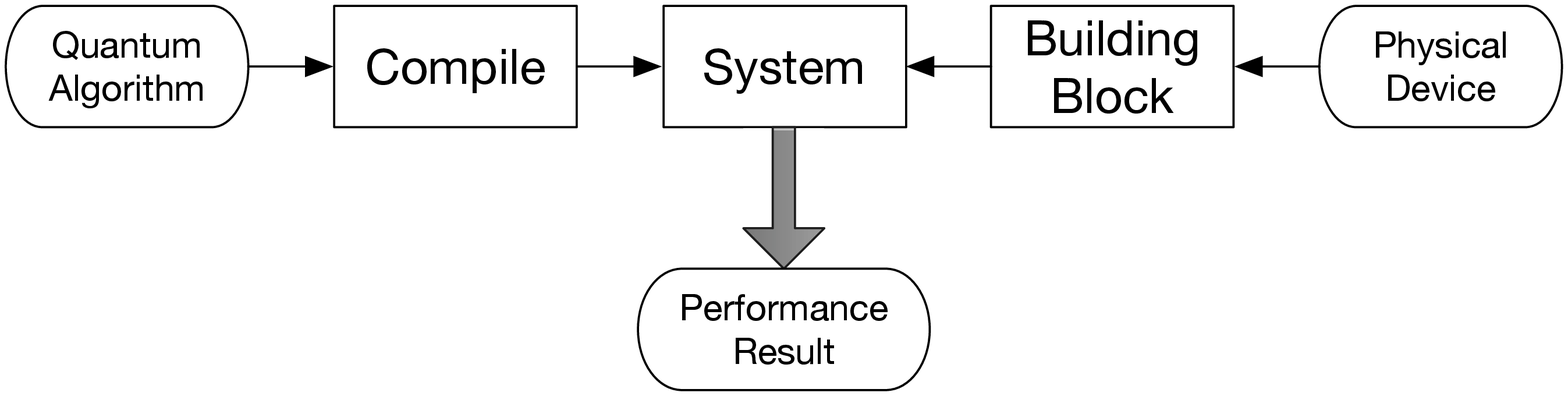, scale=0.3}
}
\subfigure[]{
	\epsfig{file=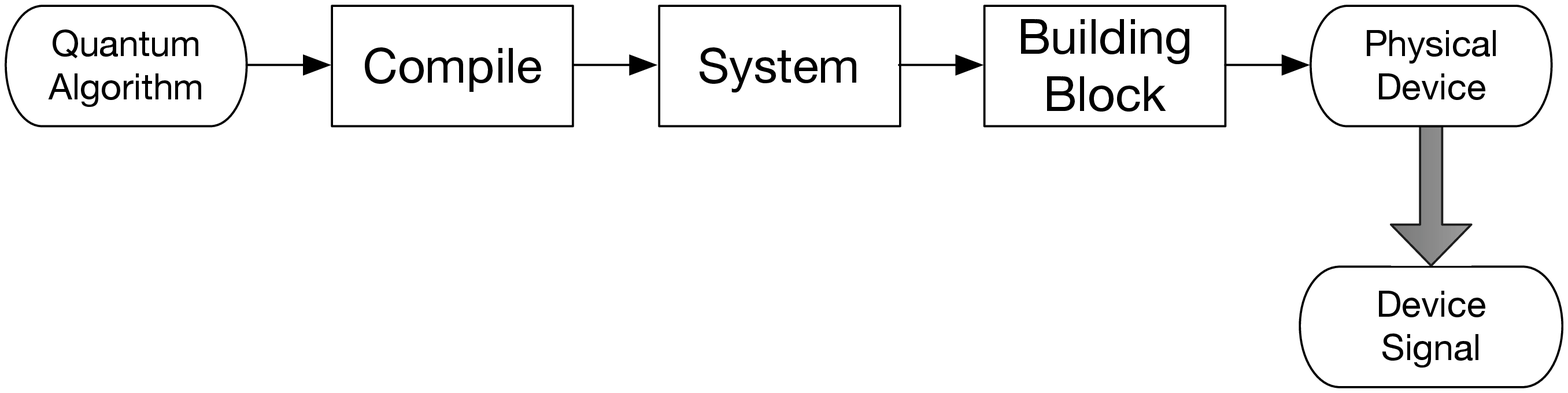, scale=0.3}
}
\caption
{
Data flow over the layers.
(a) For the performance analysis, the compile layer and the building block layer provide a quantum assembly code and the performance of logical operations to the system layer.
The system layer then performs the performance analysis during the system mapping.
(b) The platform can be utilized to perform a practical quantum computing. 
For that, all data is flowing sequentially from the most top layer to the bottom layer.
In the bottom layer control signal for physical device has to be generated for the practical computing.
}
\label{fig:data_flow_components}
\end{figure}

We describe the structure and functionality of the proposed quantum computing framework.
It deals with quantum computing programming, quantum compile, quantum computer architecture, fault-tolerant quantum computing scheme and quantum computing device.
To this end the proposed framework is composed of three functional layers: \textit{compile}, \textit{system} and \textit{building block}.
Each layer has a well-defined role and provides several options to conduct their functions.
All layers are closely related to each other.
FIG.~\ref{fig:data_flow_components} shows the data flow on the framework for performance analysis and quantum computing.

\begin{figure}[t]
\centering
\epsfig{file=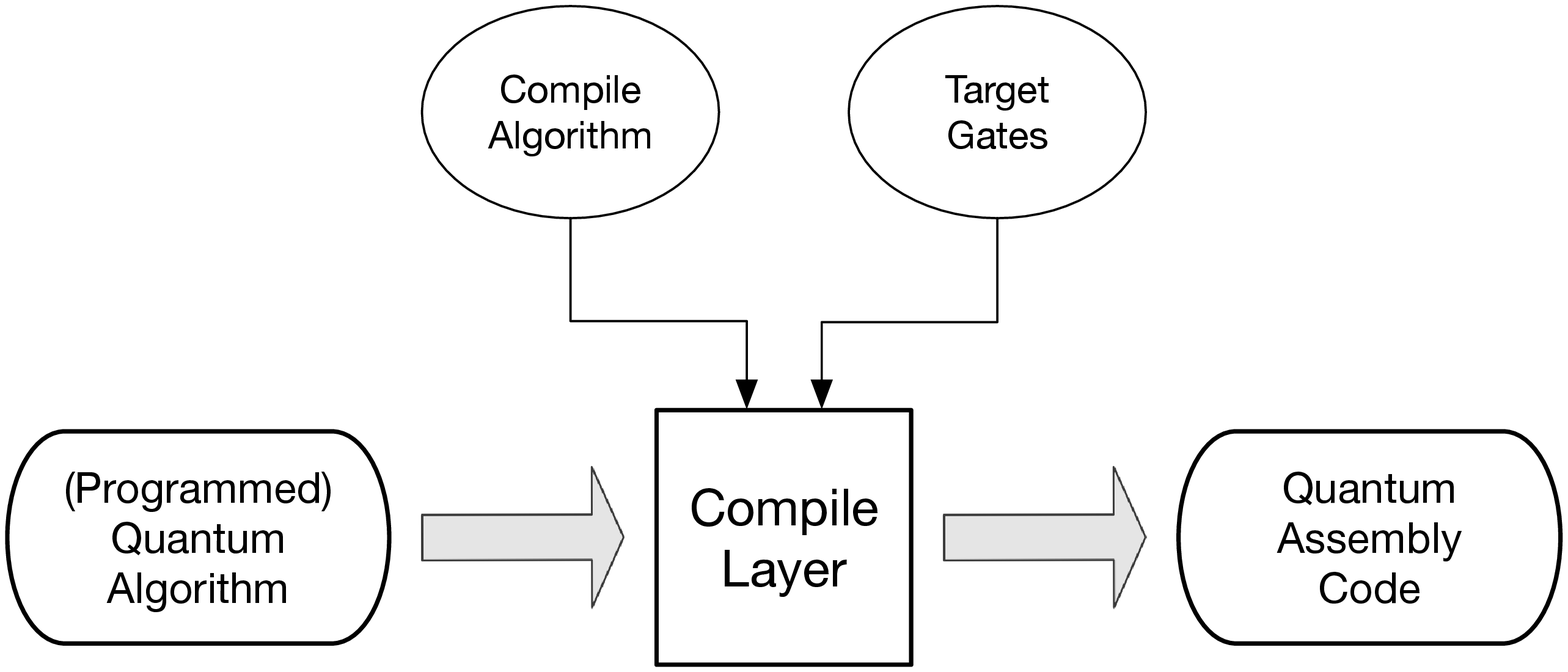, scale=0.3}
\caption{
The input and output in the compile layer.
By a compile, a programmed quantum algorithm is decomposed into a quantum assembly code.
For the compile, a compile algorithm and target gates have to be determined beforehand.
Target gates can be varied according to a quantum computing type such as a fault-tolerant quantum computing or a non-fault-tolerant (physical) computing.
The selection of target gates can also be influenced by a qubit technology.
}
\label{fig:io_compile_layer}
\end{figure}

The compile layer covers programming and compiling a quantum algorithm. 
A quantum algorithm is written in a high-level programming language, and then compiled into a sequence of quantum gates, called a \textit{quantum assembly code}.
To compile a programmed quantum algorithm, target quantum gates and compile algorithm have to be determined beforehand.
FIG.~\ref{fig:io_compile_layer} shows the input and output of the compile layer.
In the present work, we hire an open-source quantum computing compiler \textit{ScaffCC}~\cite{JavadiAbhari:2015jf,JavadiAbhari:2014eu,ScaffCCScaffCC:vm} that supports programming language \textit{Scaffold}~\cite{JavadiAbhari:2012wx}.
The details about quantum compile, quantum gates and a quantum assembly code are described in section~\ref{subsec:compile}.
Why we exploit ScaffCC (Scaffold) in this work is also discussed there.

\begin{figure}[t]
\centering
\epsfig{file=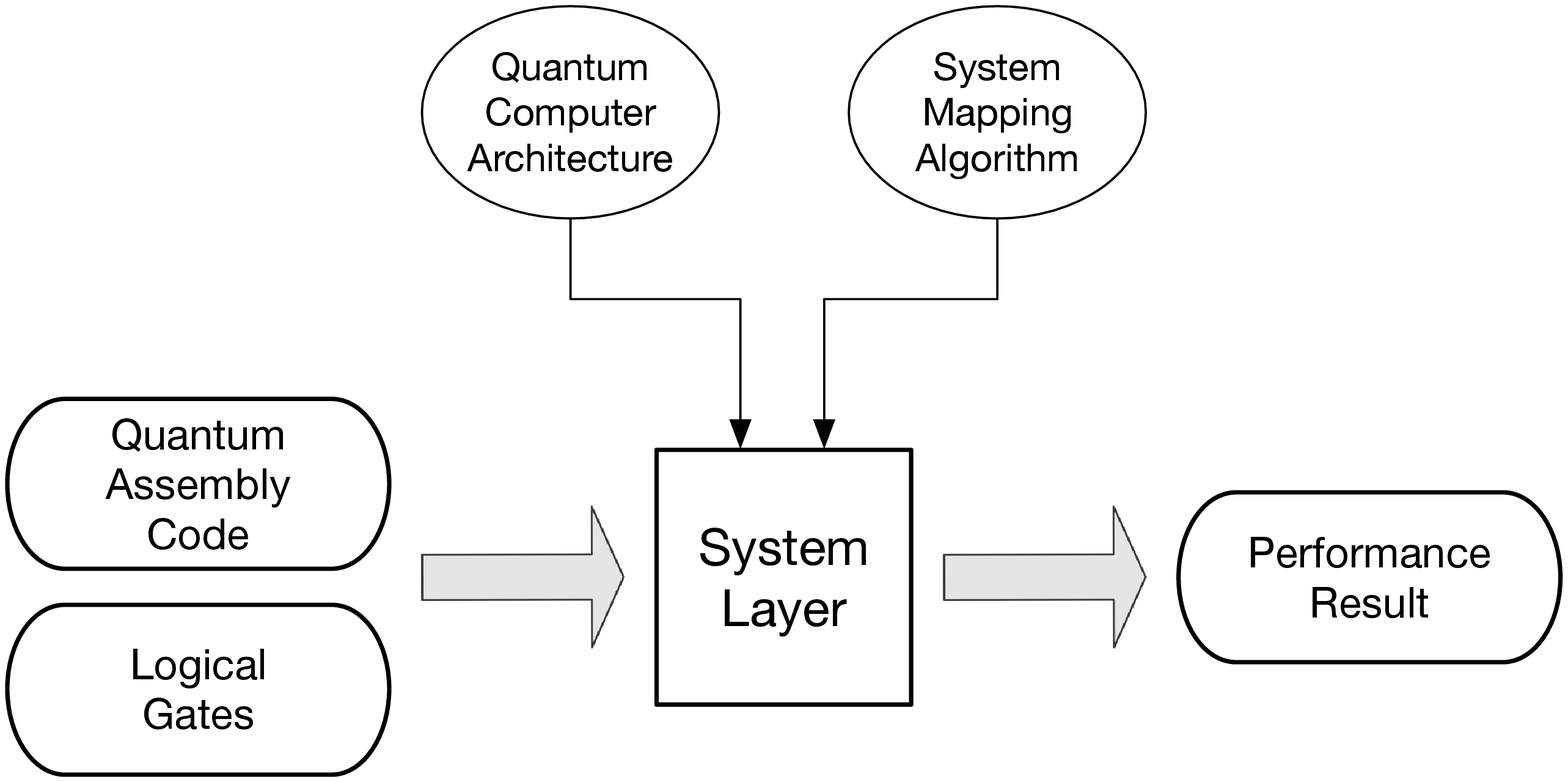, scale=0.3}
\caption{
The input and output in the system layer.
Quantum assembly code and logical gates are passed from the compile layer and the building block layer respectively.
Quantum computer architecture describes the layout of logical/physical qubits and a communication bus over qubits.
System mapping algorithm describes how to integrate a quantum algorithm (quantum assembly code) and a quantum computer architecture.
It includes the qubit placement, the gate scheduling and so on.
}
\label{fig:io_system_layer}
\end{figure}

The system layer synthesizes a quantum computing by integrating quantum algorithm, quantum computer architecture and building block (qubit/gate and quantum computing device).
It deals with everything required to run a quantum algorithm on a quantum computer system architecture.
For example, it recasts a quantum algorithm for a target quantum computer architecture via \textit{system synthesis} (also called \textit{system mapping}).
For the system synthesis, this layer first takes a quantum computing system architecture.
A qubit connectivity must be definitely defined there, and a communication bus may be included for an efficient interaction over distant qubits.
Issues related to a quantum computer architecture depend on a chosen quantum error-correcting code and a fault-tolerant quantum computing scheme.
FIG.~\ref{fig:io_system_layer} describes the input and output in the system layer.
In terms of the performance analysis, the output of the system layer is the performance analysis result, but to run a quantum algorithm this layer generates an architecture-specific description of a quantum algorithm as shown in FIG.~\ref{fig:data_flow_components}.

\begin{figure}[t]
\centering
\epsfig{file=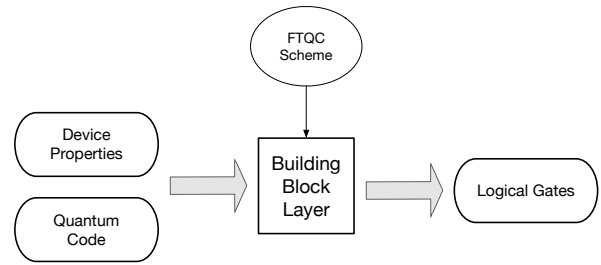, scale=0.3}
\caption{
The input and output in the building block layer.
To make logical qubits/gates, a quantum error-correcting code should be determined first.
Besides, to derive the performance of logical operations, the properties of physical device, time and fidelity, have to be considered.
Then, based on the physical device property and the fault-tolerant quantum computing scheme, logical operations with a specific performance will be generated.
}
\label{fig:io_building_block_layer}
\end{figure}

The building block layer is associated with qubits and gates of quantum algorithm.
This work basically assumes a fault-tolerant quantum computing, and therefore such qubits and gates are related to logical qubits and gates encoded in a quantum error-correcting code.
In this regards, the main functionality of this layer is to assemble physical qubits and gates to implement a logical qubit and gate.
For that, a quantum error-correcting code should be determined first, and then logical qubit and gate will be implemented according to a fault-tolerant quantum computing scheme of the chosen quantum code.
During the implementation, it is able to generate the performance of logical qubit and gate based on the properties of physical operations, time and fidelity (or \textit{error rate}).
FIG.~\ref{fig:io_building_block_layer} shows the input and output at the building block layer. 
In terms of the performance analysis, the output of the building block layer is the performance of logical quantum operations, \textit{time} and \textit{fidelity}.
Note that the proposed platform supports $[[7,1,3]]$ Steane code and a surface code.

\subsection{Configuration of Quantum Computing}\label{subsec:configuration}

We describe how to configure a quantum computing with the proposed framework.
As mentioned above, it is possible to configure a quantum computing by selectively choosing specific protocols or the properties of physical device. 
For example, you can configure a Steane code based fault-tolerant quantum computing to run Shor algorithm.
For that, through compile you generate a quantum assembly code. 
Logical building blocks are built up based on the chosen quantum error-correcting code, quantum device and the size of the quantum algorithm.
The next task is to determine a quantum computer architecture of a certain physical and/or logical qubit layout.
The architecture depends on the chosen fault-tolerant quantum computing scheme.
Note that it is also possible to choose a physical non-fault-tolerant quantum computing by assuming high-fidelity quantum computing device (see Section~\ref{sec:accurate_gates}).

\begin{table*}[t]
\caption{
List of protocols and layouts currently supported by the framework.
In the compile layer, we choose a compile type and a target gate set.
The compile type is the format of a quantum assembly code.
The compile type is closely related to the mapping type and qubit layout.
A specific qubit layout can be chosen within a selection.
In the building block layer, the scheme for a fault-tolerant quantum computing is determined.
According to the scheme, the protocol of logical operations are fixed.
\newline
}
\small
\centering
\begin{tabular}{c|c|c} \hline 
Layer & Options & Values \\ \hline
\multirow{3}{*}{Compile} & Compile Type & Structured code, Non-Structured code \\  \cline{2-3}
 & \multirow{2}{*}{Target Gate Set} & $\{X, Z, H, S (S^{\dag}), T (T^{\dag}), R_Z(\theta), CNOT\}$, \\ 
 & & $\{X, Z, H, S (S^{\dag}), T (T^{\dag}),  CNOT\}$ \\ \hline
\multirow{3}{*}{System} & Mapping Type & Structured, Non-structured \\ \cline{2-3}
 & \multirow{2}{*}{Qubit Layout} & (Non-structured) All-to-All, 1D, 2D, Arbitrary \\ 
  &   & (Structured) All-to-All, (1D, 1D), (1D, 2D), (2D, 2D) \\ \cline{2-3} \hline
Building & FTQC Scheme & Steane code, Surface code\\ \cline{2-3}
Block  & Device & Time, Fidelity \\ \hline
\end{tabular}
\label{tab:options}
\end{table*}

In Table~\ref{tab:options}, we show some options our framework currently supports.
In the table, the compile type and the mapping type (with qubit layout) completely depend on the type of a quantum assembly code, a \textit{structured} code and a \textit{non-structured} code.
A structured quantum assembly code is processed by a structured system mapping for a structured quantum computer architecture.
Similarly, a non-structured quantum assembly code is taken by a non-structured system mapping based on a simple qubit layout such as a regular 1- or 2-dimensional lattice.
The details of the structured and non-structured quantum assembly codes will be described in Section~\ref{subsec:compile}.
In Sections~\ref{sec:performance_evaluation} and~\ref{sec:performance_improvement}, we configure quantum computings with such options and analyze the performance and the quantum resource.

With the proposed framework, a quantum computing can be configured as follows.
The first step is to determine the type of a quantum computing, a fault-tolerant quantum computing or a non-fault-tolerant physical quantum computing.
In general, such determination depends on the size of a quantum algorithm and the assumption on the reliability of physical device.
The larger quantum algorithm requires the more reliable qubits and gates.
If you decided a fault-tolerant quantum computing, you need to choose a quantum error-correcting code.
Our platform supports $[[7,1,3]]$ Steane code and a surface code now.
According to a quantum algorithm and the reliability (error rate) of physical device, the concatenation level for Steane code or the code distance for a surface code will be determined.



%

The choice of the quantum computing type affects a succeeding quantum compile step. 
Exactly, target quantum gates for a compile completely depends on the chosen quantum computing type.
For example, $R_Z(\theta)$ gate for a rotational angle $\theta$ is very exploited in a quantum algorithm, but its logical version is not generally implemented in a fault-tolerant manner. 
Therefore, to realize a fault-tolerant quantum computing, you have to decompose such $R_Z(\theta)$ gate into a sequence of $H$, $S$ and $T$ gates those are fault-tolerantly implementable.
On the other hand, since the physical $R_Z(\theta)$ gate can be easily performed on physical qubits, there is no problem with quantum gates including the rotational gate for a non-fault-tolerant physical quantum computing.

The second step is to make a quantum computing program and compile it into a quantum assembly code.
As mentioned above, for the compile, you have to exploit target quantum gates determined in the previous step.
The proposed framework exploits open-source programming language \textit{Scaffold} and compiler \textit{ScaffCC}.
You can see how to make a Scaffold program in Ref.~\cite{JavadiAbhari:2012wx} and how to use ScaffCC compiler in Ref.~\cite{JavadiAbhari:2015jf,ScaffCCScaffCC:vm}.
In section~\ref{subsec:compile}, we show a simple example of a Scaffold program and the associated quantum assembly code. 
Note that ScaffCC decomposes an arbitrary one-qubit gate into a sequence of $H$, $S$ and $T$ by exploiting \textit{gridsynth}~\cite{Ross:2016uz} or \textit{sqct}~\cite{Kliuchnikov:2013tr}.

The third step is to choose a quantum computing architecture, in particular a qubit array. 
The proposed platform takes not only a simple regular 1- or 2-dimensional lattice, but also a hierarchically structured qubit array.
A communication bus also should be considered to make an efficient interaction over distant qubits.
FIG.~\ref{fig:layout_example} shows an example of a hierarchically structured quantum computing architectures.
In case of Steane code quantum computing, the structure of a qubit array seriously affects a quantum computing due to the limited local qubit interaction.
We will show such limitation raises highly nontrivial temporal overhead in section~\ref{sec:performance_local_qec}.
On the other hand, surface code quantum computing scheme is fundamentally established based on local qubit interaction on the 2-dimensional lattice.
FIG.~\ref{fig:surface_code_architecture} shows a quantum computing architecture for a surface code quantum computing.

If the configuration is completed, we can perform the system synthesis.
During the system synthesis, the quantum algorithm is reformulated for the target quantum computer architecture. 
As will be discussed later, from the system synthesis, the expected performance (circuit depth, execution time, fidelity, KQ, and so on) and the required quantum resource (qubits and gates) of a quantum computing are evaluated.

\subsection{Analysis of Quantum Computing}\label{subsection:analysis}

We briefly mention what performance are evaluated by the framework, but the detailed analysis method will be described in section~\ref{sec:performance_metric}. 
The framework first inspects the quantities of qubits and gates.
Those figures are usually analyzed by a quantum compile without considering a quantum computer architecture~\cite{Smith:2014ux,JavadiAbhari:2015jf}.
But, our framework examines such quantities with the consideration for a quantum computer system architecture. 
In the system synthesis, a fault-tolerant quantum computing scheme in particular a magic state factory is taken into consideration.
Therefore, it is possible to estimate the temporal and spatial overhead caused by factors those are veiled in a quantum algorithm, and therefore we believe our estimation nearly coincides with the resource to perform a real quantum computing.

The framework examines the expected quantum computing execution time (circuit depth, fidelity, KQ and so on) of a quantum algorithm based on quantum computer architecture, fault-tolerant protocol and physical device.
By applying the properties of physical device and fault-tolerant protocol, we deduce the performance of logical gates and quantum error correction.
Besides, by conducting a system integration, we are able to obtain the single-round execution time of a quantum algorithm by applying the performance of logical gates and error correction above.
Our framework goes further for more detailed analysis.
During the system integration, as mentioned above, the fidelity of a quantum computing can be calculated. 
By reflecting the fidelity, it is possible to estimate the execution time for a quantum computing achieving a fidelity 100$\%$.
In doing so, we fairly compare fault-tolerant quantum computing and non-fault-tolerant quantum computing.
The quantity of qubits is also based on this performance criterion.

Besides the above-mentioned, the framework can generate various performance data.
Based on the data, we can estimate the temporal and spatial overhead of a quantum computing.
For example, as mentioned above limited local interaction between nearest neighbor qubits sometimes requires additional qubit movements to perform 2-qubit CNOT gate.
The quantity of such qubit movements corresponds to the temporal overhead.
The platform evaluates such temporal overhead as a ration of SWAP gates to total quantum gates.
As will be shown in section~\ref{sec:performance_local_qec}, a quantum computing requires highly nontrivial temporal overhead.

\section{Proposed Quantum Computing Framework}\label{sec:proposed_framework}
\subsection{Compile Layer}\label{subsec:compile}

A quantum compile is a process that decomposes a quantum algorithm into a sequence of quantum gates.
Here a quantum algorithm is in entirely programmed form by using a high-level abstract programming language.
Recently, several research groups have developed programming environments for a quantum computing by modifying conventional classical programming languages such as python and C/C++~\cite{Steiger:2018to,JavadiAbhari:2012wx,Microsoft:wz,Ying:2017exa,Green:2013wb,Green:2013gi}.

It is well known that an arbitrary quantum algorithm can be decomposed into a combination of 1-qubit rotational gates and 2-qubit $CNOT$ gate~\cite{Nielsen:2000ga}.
The target quantum gates for a compile can be varied according to situation.
For example, the set of $H$, $T$, and $CNOT$ is de facto standard for a universal fault-tolerant quantum computing.
But, to reduce the complexity in compile or to provide a flexibility to a programmer, one usually add more quantum gates to target gates.
Furthermore, according to specific quantum mechanical system, physically implementable quantum gates slightly differ~\cite{Nielsen:2000ga,Lin:2014il}. 
In this work, we utilize two sets of quantum gates $\{X, Z, H, S (S^{\dag}), T (T^{\dag}), R_Z(\theta), CNOT\}$ and $\{X, Z, H, S (S^{\dag}), T (T^{\dag}), CNOT\}$.
The difference between both is $R_Z(\theta)$ gate.
As mentioned before, since the rotational gate for an arbitrary rotational angle $\theta$ can not be implemented in the fault-tolerant manner, the first set is exploited for a physical quantum computing and the second is used for a fault-tolerant quantum computing.

The output of the quantum compile, the sequence of quantum gates, is called a \textit{quantum assembly code}.
A quantum assembly code is a list of quantum instructions that is a combination of a quantum gate and its target qubit(s).
It is an intermediate representation of a quantum algorithm between a mathematical description and a physical machine instruction description~\cite{Cross:2017ud,JavadiAbhari:2015jf,Svore:2006iw}.
There is not any standard for a quantum assembly code, and therefore a specific representation and a structure of which slightly differ according to literature.
For example, a quantum instruction to apply a Hadamard gate to a qubit $q$ is represented as ``$hadamard$ $q$" or ``$H$($q$)".
Besides, a certain quantum assembly code has its own special structure.
For example, Open QASM by IBM~\cite{Cross:2017ud} provides a conditional statement ``\textit{if... then... else}" usually supported by conventional programming languages.

\begin{figure}[t]
\centering
\subfigure[]{
	\epsfig{file=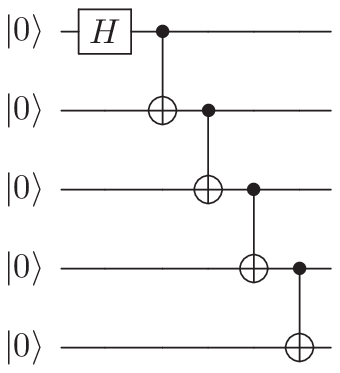, scale=0.8}
}
\subfigure[]{
	\epsfig{file=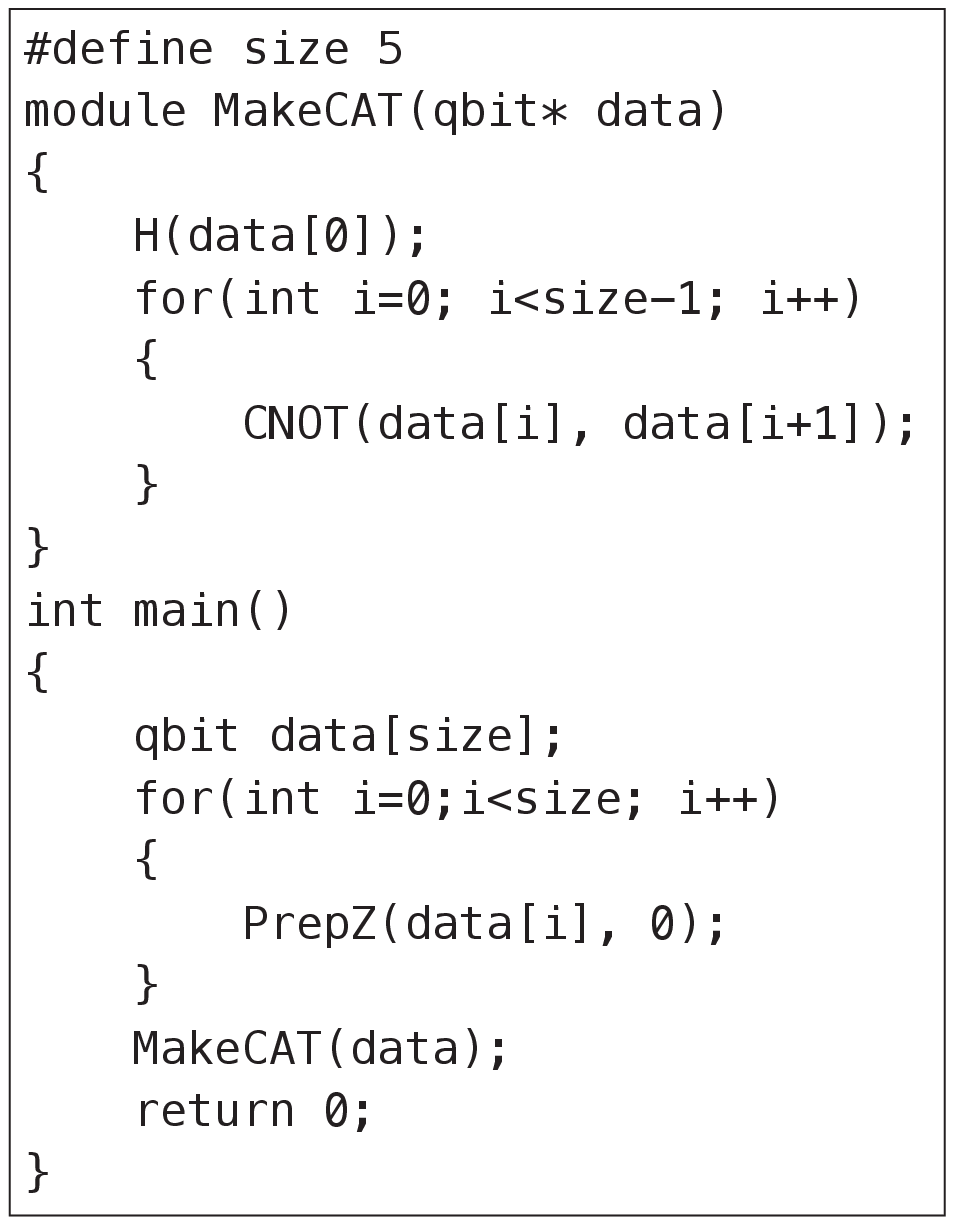, scale=0.4}	
}
\caption{
	An example of a Scaffold program to implement a $5$-qubit CAT state.
	(a) A quantum circuit and (b) its Scaffold program.
	The module \textit{MakeCAT} makes the CAT state.
}
\label{fig:scaffold_code}
\end{figure}

In the present work, we hire open-source quantum compiler \textit{ScaffCC}~\cite{JavadiAbhari:2015jf,ScaffCCScaffCC:vm}. 
It compiles a quantum computing program written in quantum computing programing language \textit{Scaffold}~\cite{JavadiAbhari:2012wx}.
A Scaffold program consists of one \textit{main} module and multiple sub-modules (see FIG.~\ref{fig:scaffold_code}).
A module generally seems like a function or a procedure of conventional programming languages such as $C/C++$, Python and so on.
The module is composed of instructions calling quantum gates and/or other modules.
The execution of a Scaffold program begins with the first instruction of the main module, and terminates by conducting the last instruction of the same module.
During the execution of the main module, other sub-modules may be called.

\begin{figure}[t]
\centering
\epsfig{file=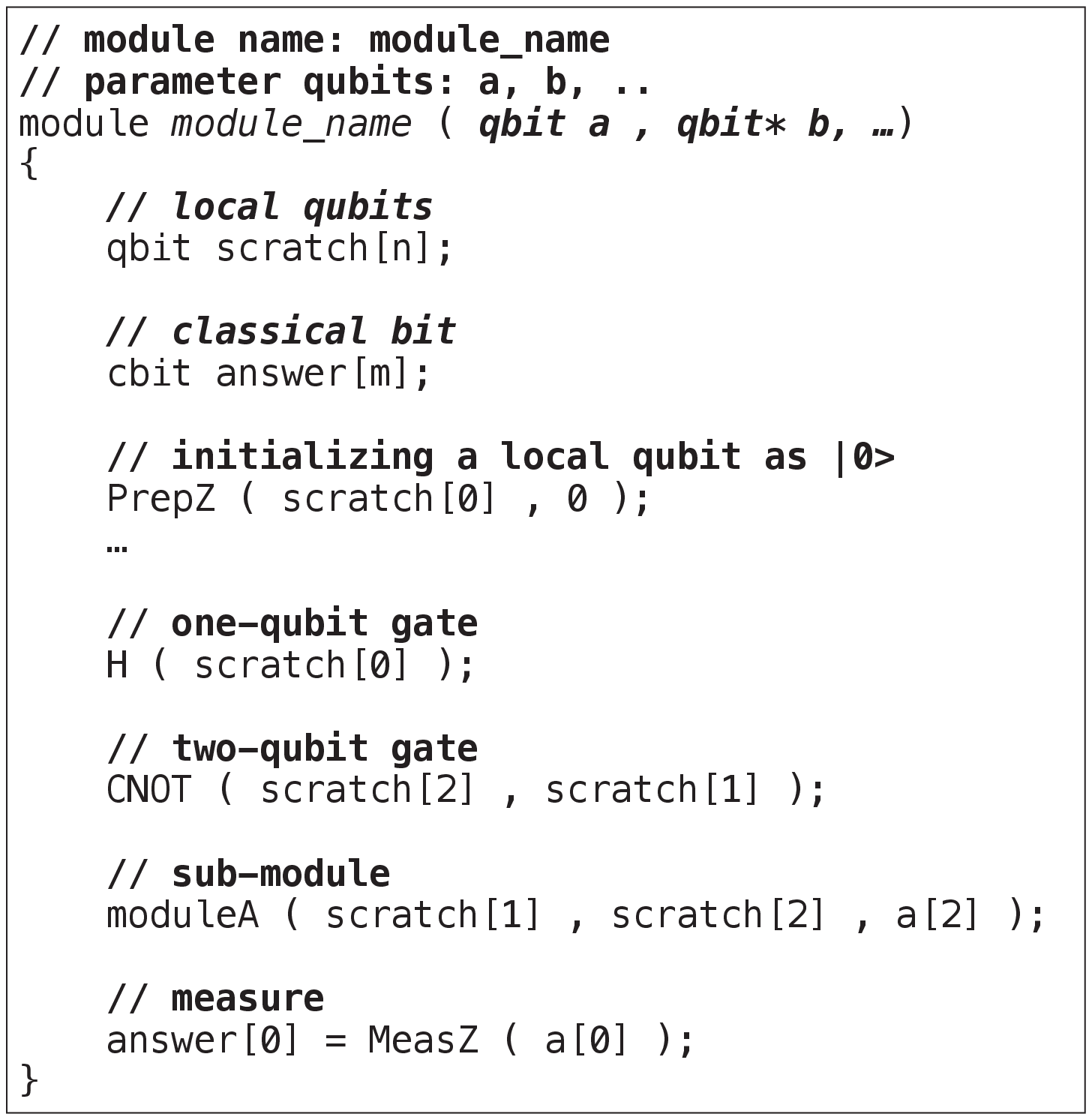, scale=0.45}
\caption{
An example of a module. 
Parameter qubits passed from external modules are clearly specified at the beginning.
A module is defined by the preparation of local qubits and classical bits, calling quantum gates and other sub-modules, and measurement of local qubits.
}
\label{fig:module_model}
\end{figure}

Previously, we mentioned that some quantum assembly codes have unique feature in their structure.
So does the output of ScaffCC.
The compiler generates a hierarchically structured quantum assembly code, in which a quantum algorithm is composed of multiple modules.
A module consists of performing quantum gates and/or other modules. 
In the previous paragraph, we also mentioned that a Scaffold program consists of modules.
To avoid any ambiguity, we need to distinguish both.
A module in a Scaffold program is completely defined and written by a programmer, and through a compile it is converted to a module in a quantum assembly code.
Therefore, both are technically identical.
FIG.~\ref{fig:module_model} shows an example of a module in a quantum assembly code.

\begin{figure}[t]
\centering
\epsfig{file=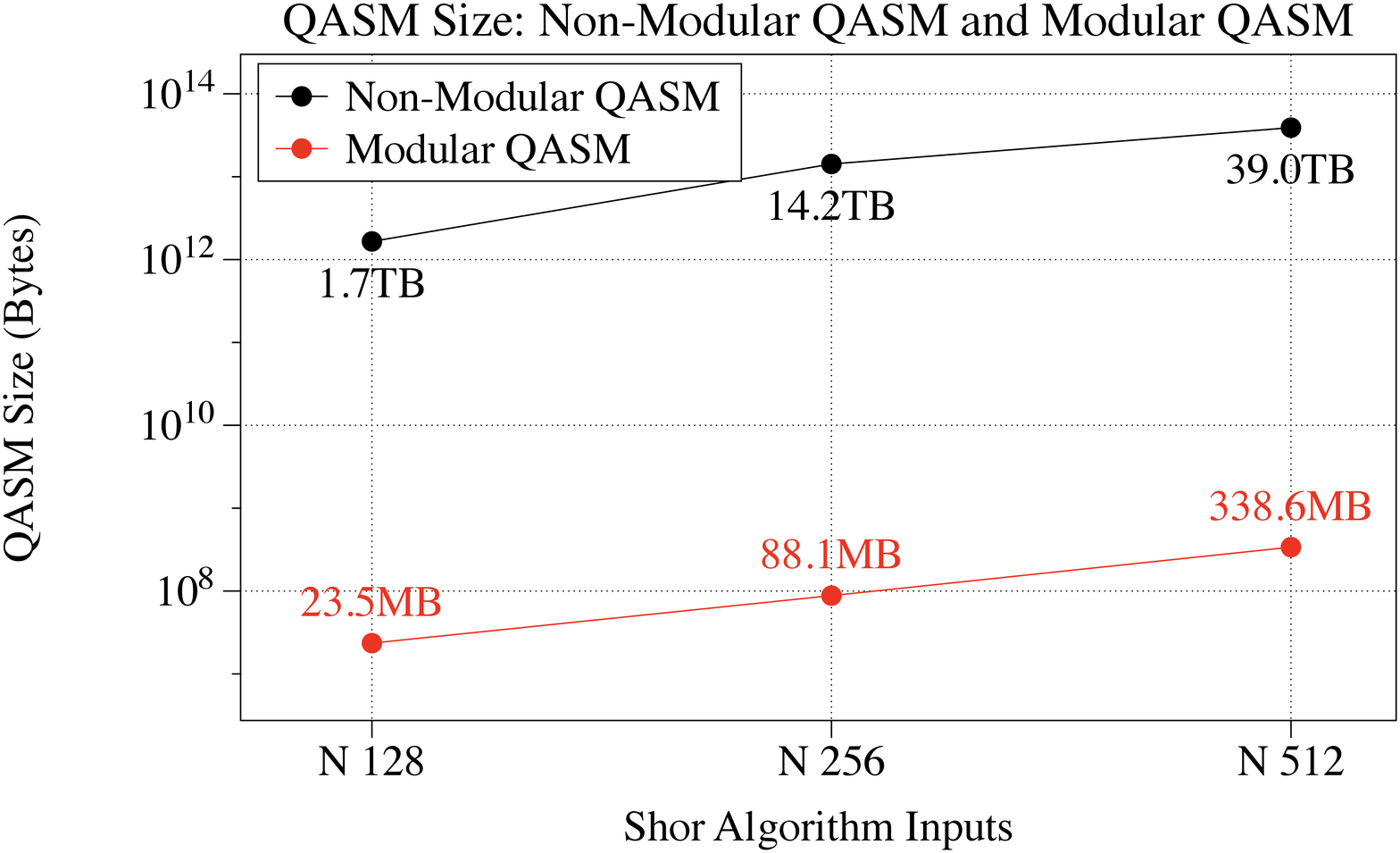, scale=0.23}
\caption{
The size of the quantum assembly codes for Shor algorithm $n=128$, $256$, and $512$.
Both codes are generated by ScaffCC.
As the input increases, the size of quantum assembly code in the non-structured format increases over dozens TB.
}
\label{fig:qasm_size_comparison}
\end{figure}

We need to say why we exploit ScaffCC in this work.
As mentioned above, a quantum assembly code is a list of quantum instructions.
As the size of a quantum algorithm increases, the size of a quantum assembly code nontrivially increases.
It completely follows the complexity of a quantum algorithm.
Obviously, the size of meaningful quantum algorithm in reality is beyond the capability of conventional super-computing.
In other words, the size of a quantum assembly code for our interested algorithm will be very huge, and such scalability will cause a practical problem in classical control of a quantum computing.
For example, the size of a quantum assembly code of Shor algorithm to factorize 512 bit integer is around 39 TB (see FIG.~\ref{fig:qasm_size_comparison}).
Therefore, we have had trouble in generating and managing such a huge code with a classical computing system.
We could not even attempt to generate more larger-sized quantum algorithm than the algorithm above due to the lack of classical storage and memory.

On the other hand, it is believed that the hierarchy provided by ScaffCC can suppress the scalability.
For example, to perform a composite quantum operation of $N$ gates as much as $M$ times, a normal (non-structured) quantum assembly code requires $MN$ quantum instructions ($\# gates \times \# iteration$).
However, the hierarchical assembly code requires only $M+N$ instructions ($\# gates + \# iteration$) by defining the operation as a module.
By doing this, the hierarchical quantum assembly code is much smaller than the non-structured code as shown in FIG.~\ref{fig:qasm_size_comparison}.
To the best of our knowledge, only ScaffCC supports such hierarchically structured quantum assembly code.
This is the main reason whey we exploit ScaffCC in the proposed platform.

\begin{figure}[t]
\centering
\subfigure[]{
	\epsfig{file=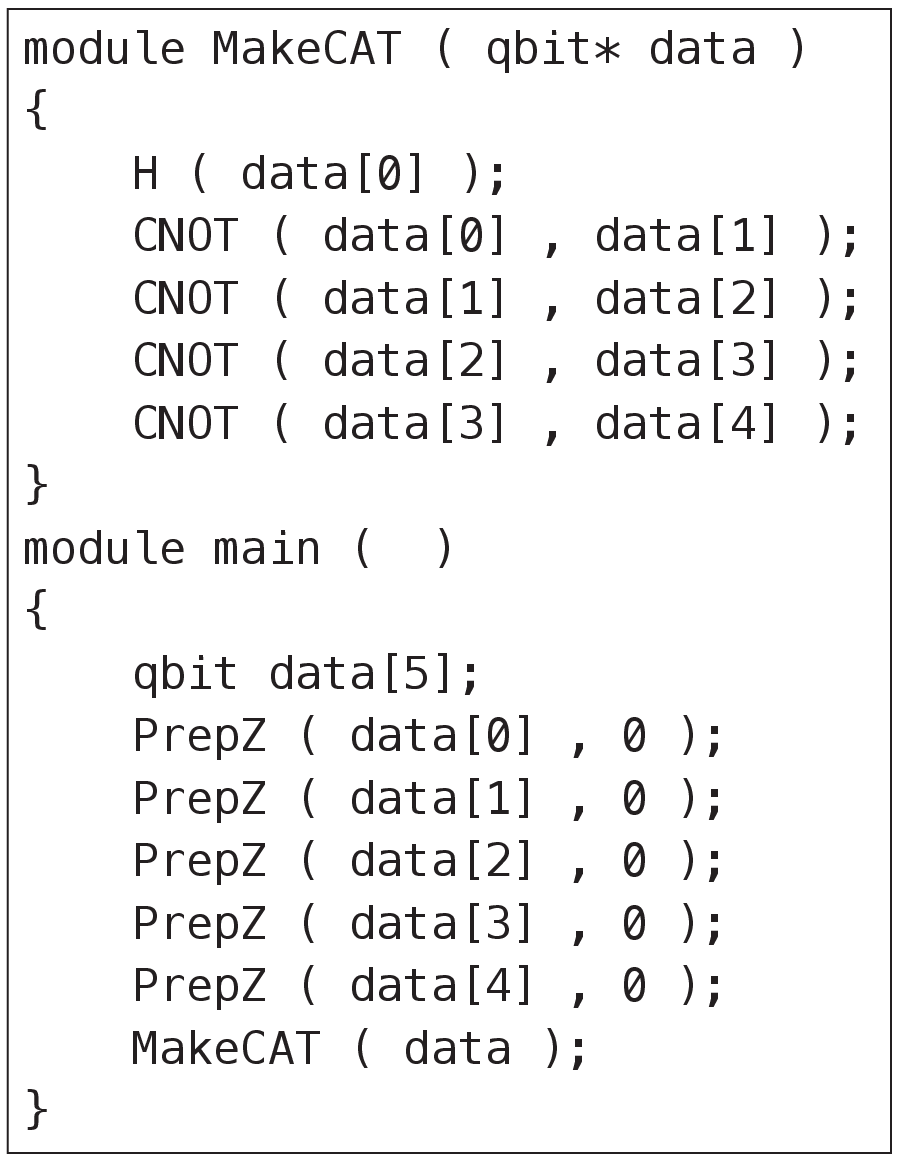, scale=0.4}
}
\subfigure[]{
	\epsfig{file=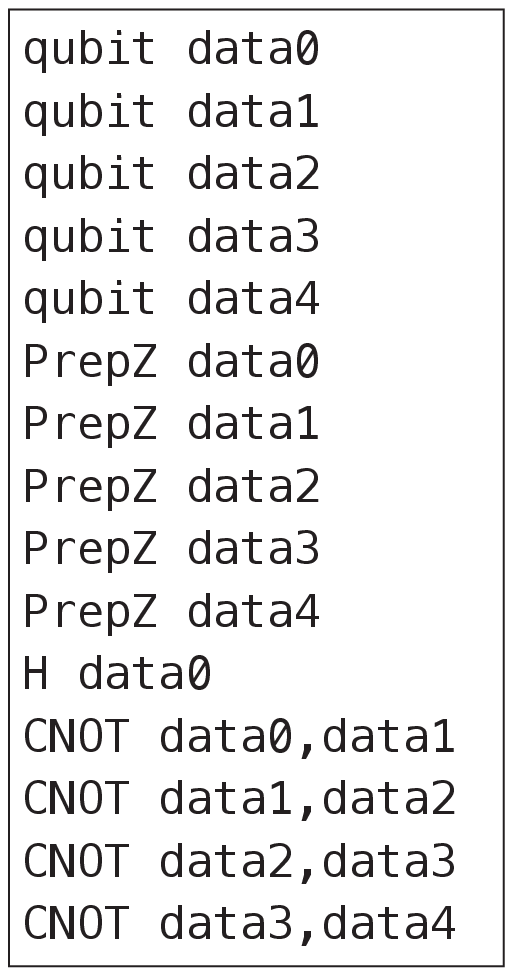, scale=0.45}
}
\caption
{
	Quantum assembly codes to generate a $5$-qubit CAT state.
	(a) A structured code and (b) a non-structured code.
}
\label{fig:cat_qasm}
\end{figure}

To conclude this section, we show a simple example of a Scaffold program to make a $5$-qubit CAT state $\frac{1}{\sqrt{2}}(|0\rangle^{\otimes 5} + |1\rangle^{\otimes 5})$ and a corresponding quantum assembly code.
Readers can see how to make a Scaffold program in Ref.~\cite{JavadiAbhari:2012wx} and how to use ScaffCC compiler in Ref.~\cite{ScaffCCScaffCC:vm}.
FIG.~\ref{fig:scaffold_code} shows a quantum circuit to implement a $5$-qubit CAT state and a corresponding Scaffold program. 
A structured and a non-structured quantum assembly code are respectively shown in FIG.~\ref{fig:cat_qasm}.

\subsection{System Layer}\label{subsec:system}

A quantum algorithm (quantum assembly code) is a logic of how to solve a given problem.
It is based on ideal physical situation where noiseless physical gates and arbitrary long qubit interaction are allowed.
In other words, a quantum algorithm is developed without considering any physical implementation.

\begin{figure}[t]
\centering
\subfigure[]{
	\epsfig{file=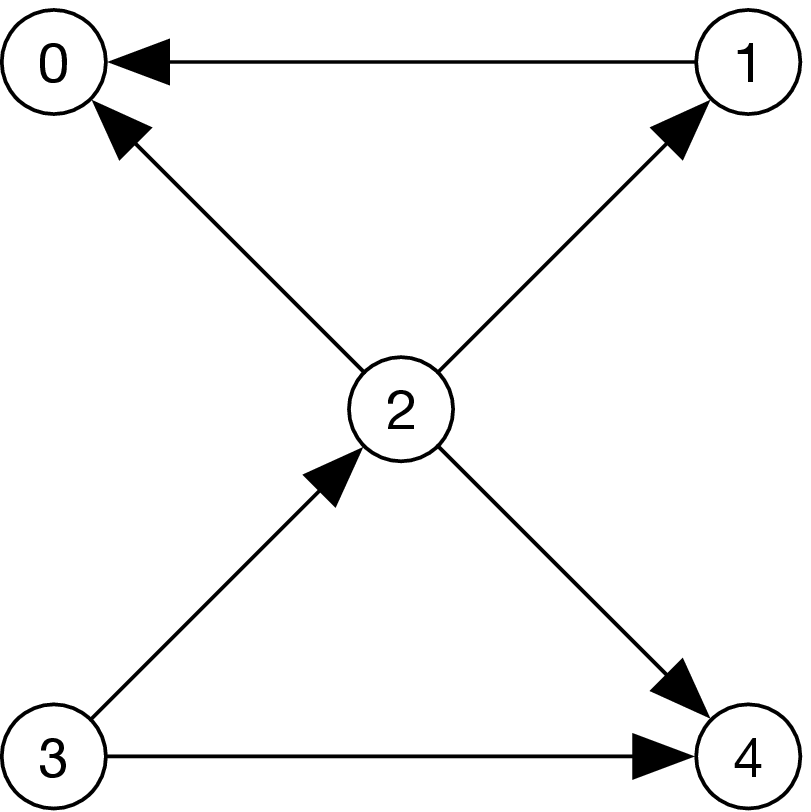, scale=0.4}
}
\subfigure[]{
	\epsfig{file=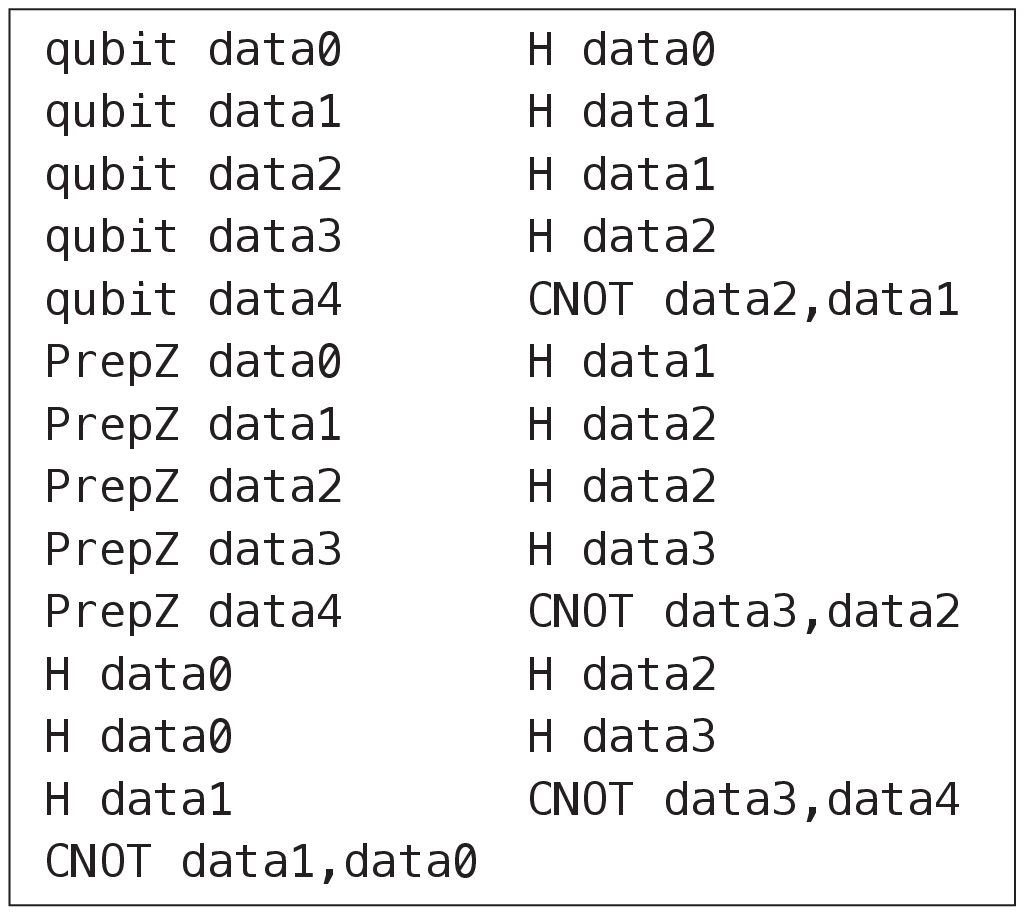, scale=0.35}
}
\caption
{
(a) The qubit layout of IBM QX4 device~\cite{IBM:Iidq8v80}. 
A node indicates a qubit, and an edge with a direction implies that the application of a controlled-CNOT gate is possible, where the control qubit and the target qubit are the root and end of the arrow. 
Therefore, as you see no bi-directional CNOT is allowed on the IBM QX4.
(b) The recasted assembly code from FIG. 7 (b). 
Since the instruction ``\textit{CNOT data0,data1}`` is not allowed directly on the IBM QX4, Hadamard gates, ``\textit{H data1}" and ``\textit{H data2}", have to add before and after of the instruction.
Note that the node index $k$ indicates the qubit data$k$.
We have not cancel out repetitive Hadamard gates.
By cancelling out those gates, the circuit depth can be reduced from 12 to 9.
}
\label{fig:mapping_simple_example}
\end{figure}

On the other hand, a quantum computer where a quantum algorithm is executed has a certain logical and physical architecture such as a qubit layout. 
Therefore, to run a quantum algorithm on such quantum computer, we need to reformulate the algorithm to be compatible with the given quantum computer architecture.
For example, real quantum computing devices in IBM Quantum Experience~\cite{IBM:Iidq8v80} have very limited qubit layout and allow only one-directional CNOT.
Therefore, the quantum assembly codes shown in FIG.~\ref{fig:cat_qasm} can not be directly executed on the IBM QX4 device (see FIG.~\ref{fig:mapping_simple_example} (a)) because the codes include unallowable CNOT gates. 
Therefore, for the execution, we have to recast a quantum assembly code for IBM QX4. 
In FIG.~\ref{fig:mapping_simple_example} (b), we show the recasted (non-structured) quantum assembly code for the device.
This is the motivation of a quantum computing system mapping.

The principle of a quantum computing system mapping is very simple, 1) \textit{set up} a quantum computer architecture and 2) \textit{recast} a quantum assembly code for the architecture.
In what follows, we first describe a quantum computer architecture, and then show how to actualize a quantum algorithm on the target architecture.

\subsubsection{Quantum Computer Architecture}
We discuss a hierarchically structured quantum computer architecture for the proposed framework.
In general, there is no restriction to the architecture.
In other literature, a regular 1- or 2-dimensional lattice is usually exploited.
But, in this work we assume that it is hierarchically structured.
A quantum computer is composed of several computing regions called a module and a communication bus connecting to the modules.
By assuming such structured architecture, the system mapping with a hierarchically structured quantum assembly code can be done efficiently.

%
%
%
%

\begin{figure}[t]
\centering
\subfigure[]{
	\epsfig{file=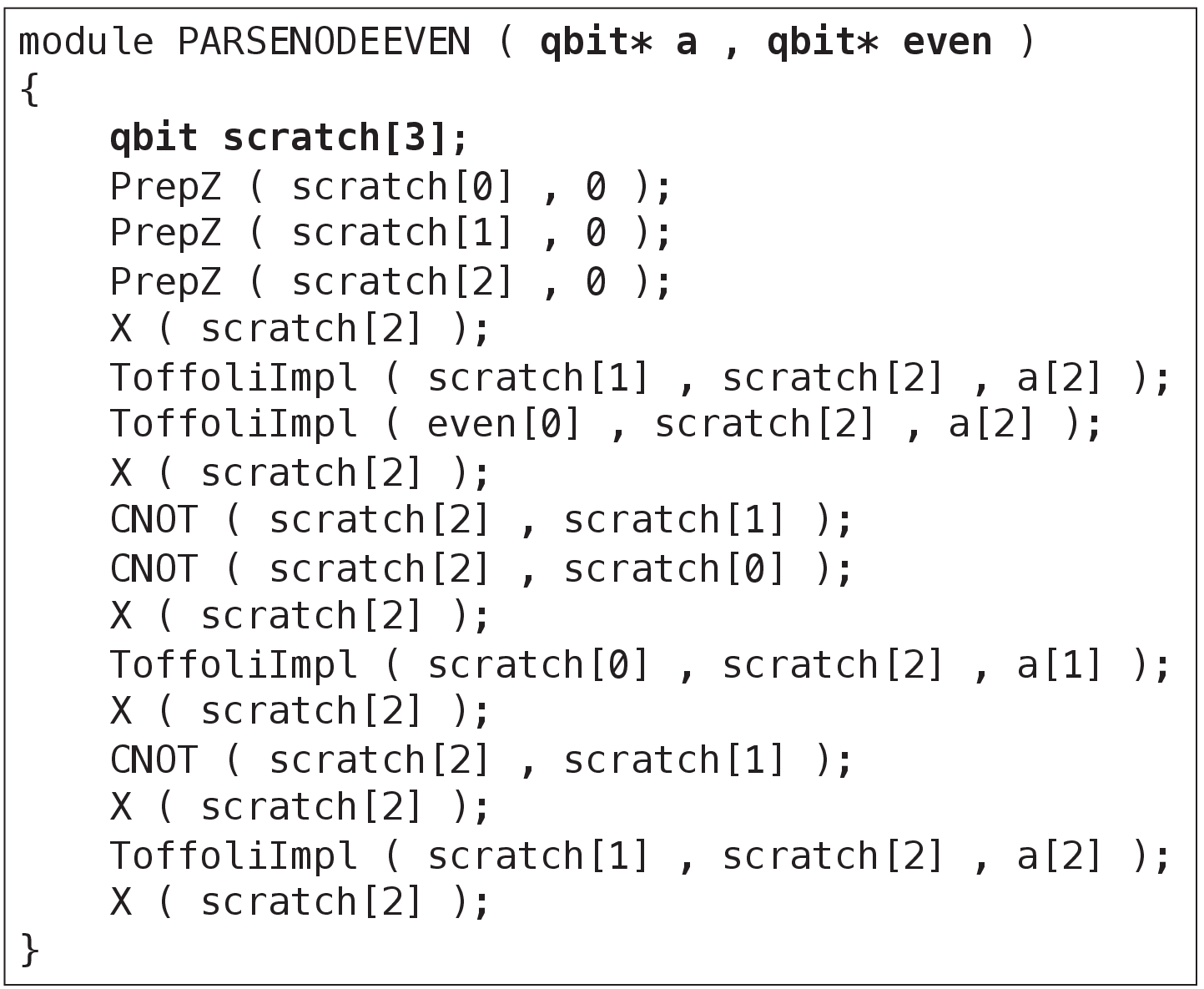, scale=0.35}
}
\subfigure[]{
	\epsfig{file=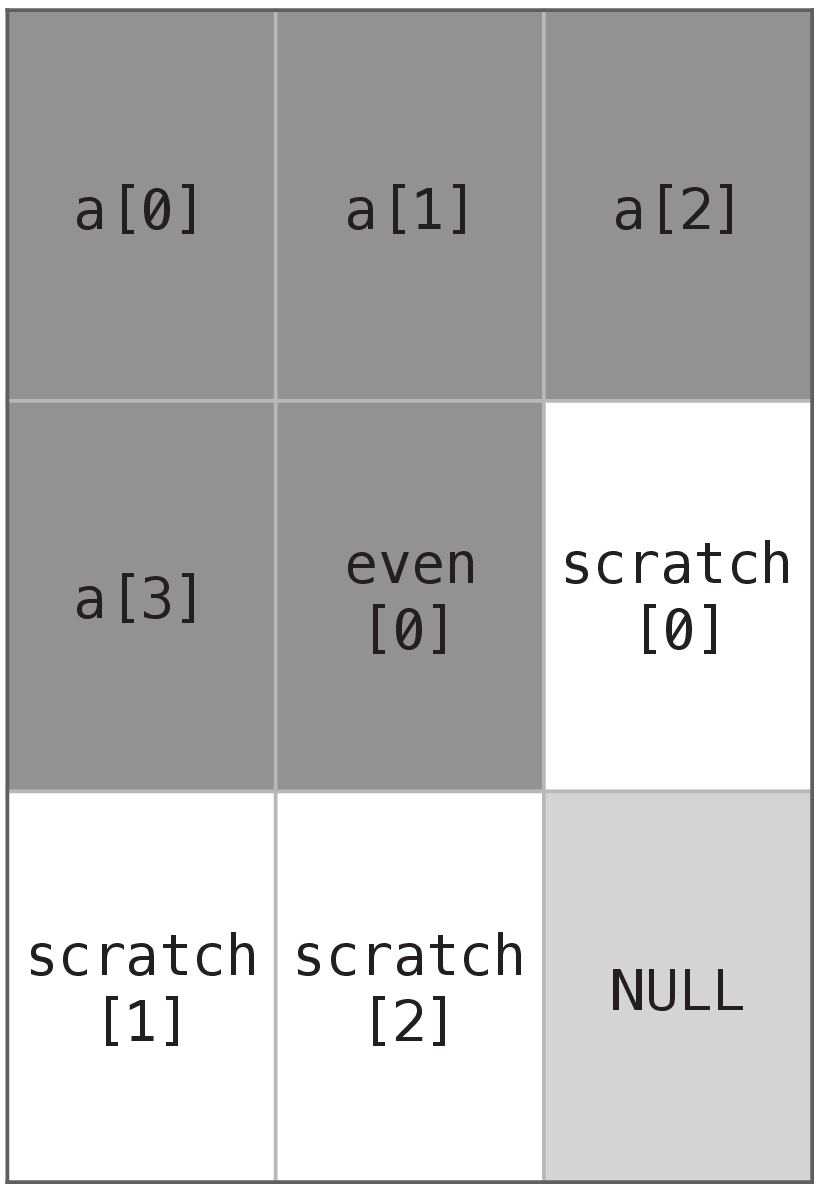, scale=0.35}
}
\caption
{
(a) An example of a module in a quantum assembly code and (b) the associated computing region on a quantum computer architecture.
In (a), the qubits \textit{a} and \textit{even} are parameter qubits passed from other modules, and the qubit \textit{scratch} is a local qubit locally used within the module.
In (b), the dark grey cells are for the parameter qubits, the white cells are for the local qubits and the light grey cells are just empty space or null qubits not working anything.
While the size of the parameter qubits \textit{a} and \textit{even} are not specified in the module definition, it can be determined by tracing all modules that calls the module. 
}
\label{fig:example_module_layout}
\end{figure}

A computing region is completely associated with a module in a quantum assembly code.
It consists of multiple cells for logical (or physical) qubits described in the module. 
Some cells are allocated for parameter qubits passed from other modules, and others are allocated for local qubits which are temporarily used within a module.
Additional space may be sometimes required to form the rectangular shape of a module.
FIG.~\ref{fig:example_module_layout} shows an example of a module in a quantum assembly code and its associated computing region.

\begin{figure}[t]
\centering
\subfigure[]{
	\epsfig{file=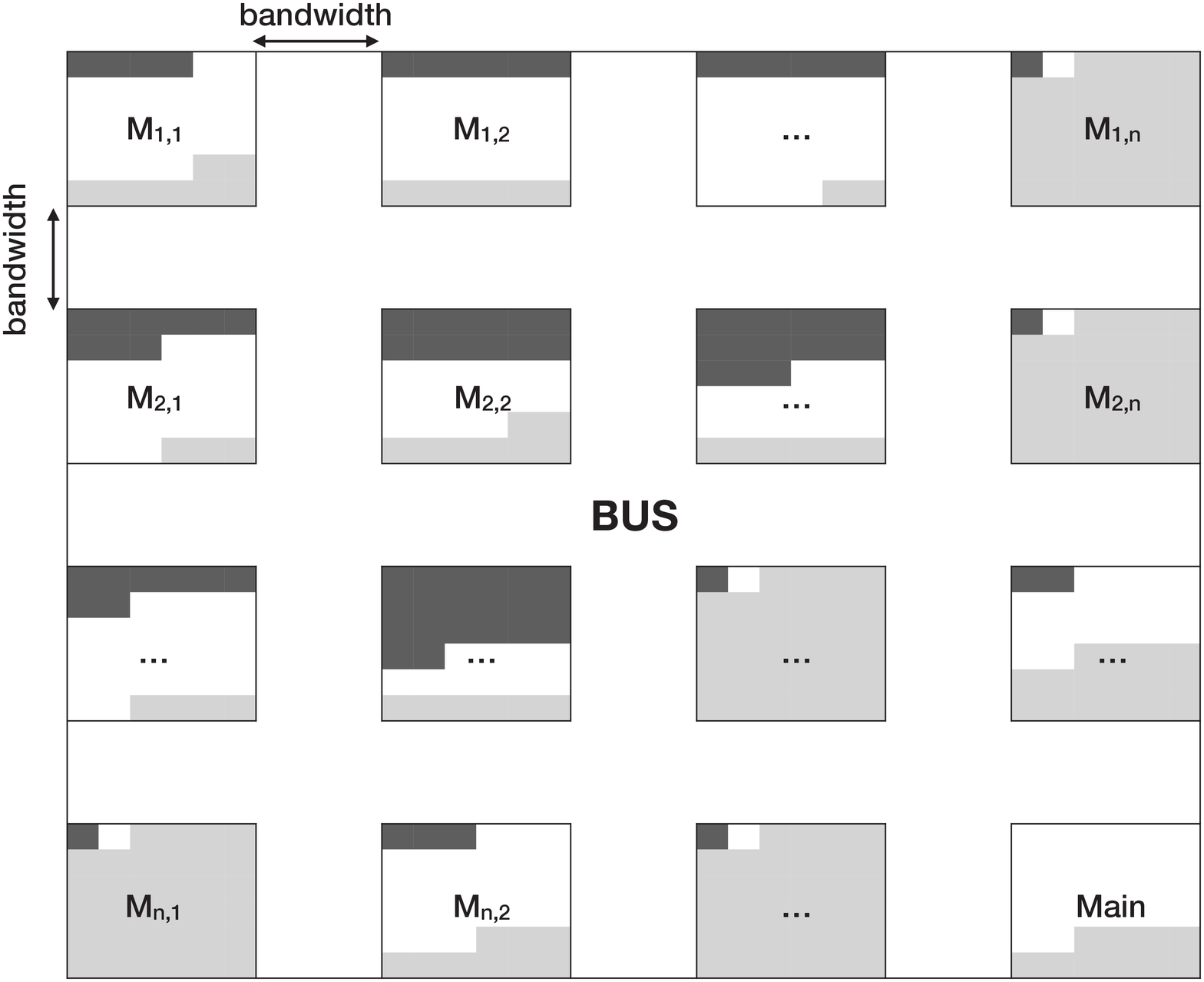, scale=0.25}
}
\subfigure[]{
	\epsfig{file=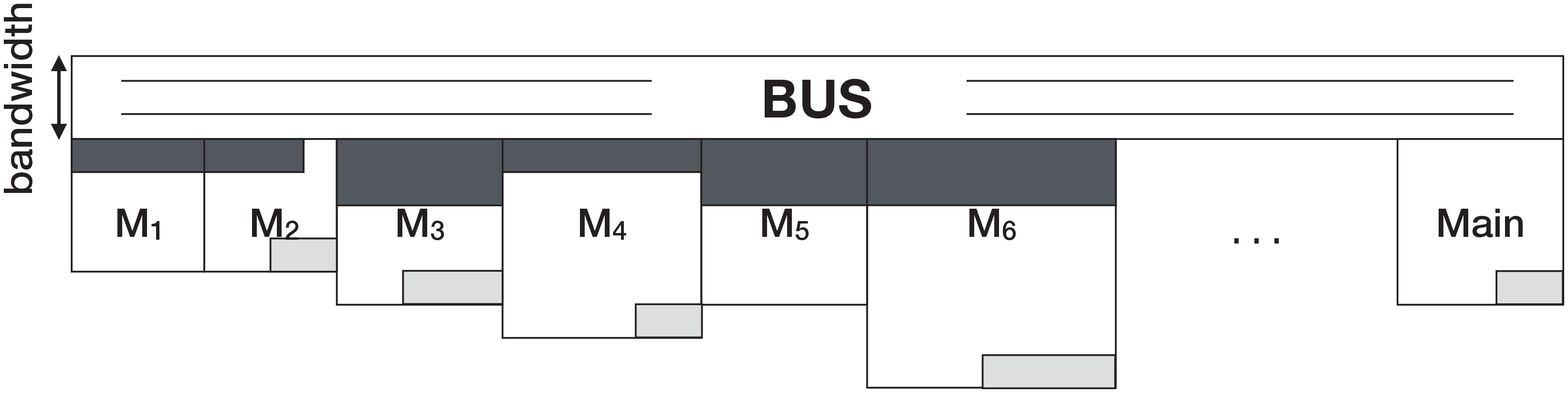, scale=0.25}
}
\caption{
An example of a proposed quantum computer architecture.
(a) 2D global layout and 2D local layout.
(b) 1D global layout and 2D local layout.
tLogical qubits of different color play different role in a module; parameter qubits (dark grey), local qubits (white) and dummy qubits (light grey).
}
\label{fig:layout_example}
\end{figure}

FIG.~\ref{fig:layout_example} shows examples of the above-mentioned quantum computer architecture.
The box labelled $M_{i,j}$ ($M_i$) indicates a module (computing region).
We call the arrangement of modules a \textit{global} layout and the arrangement of qubits within a module a \textit{local} layout.
All modules communicate with each other via a communication bus.
In the figure, the bus is depicted as a white space outside of modules.
We will discuss the bandwidth of the communication bus in Section~\ref{sec:performance_metric}.

Qubit that resides inside a module supports universal quantum operation.
The logical qubit is composed of data qubits for holding data and ancilla qubits for error correction and logical operations.
On the other hand, qubits for a communication bus only perform error correction and logical Clifford operations.
Therefore, the composition of logical qubits for a communication bus can be differ from that for modules according to a quantum error-correcting code and a fault-tolerant quantum computing scheme.


\subsubsection{System Mapping}
FIG.~\ref{fig:data_flow_components} shows that all data are collected in the system layer and the performance of a quantum computing is evaluated there.
In this section, we will describe the system mapping in terms of the gate reformulation for the target quantum computer structure and the performance evaluation.
A specific mapping process definitely depends on the type of quantum instructions.
Quantum instructions in the hierarchically structured quantum assembly code are classified into three types: 1-, 2-qubit gate and module.

The set of 1-qubit gates includes $X$, $Z$, $H$, $S$ ($S^{\dag}$), $T$ ($T^{\dag}$), $R_Z(\theta)$ and a preparation and a measurement in the Z basis.
The mapping process for such gates is straightforward and can be done independently. 
Suppose that a Hadamard gate is applied to a qubit $q$.
If the application of a certain quantum operation to the qubit was scheduled previously, the application of Hadamard gate will be performed after the previous operation.
If the previous operation was over at time $t(q)$, then Hadamard operation will start at $t(q)$ and finish at $t(q)+H_t$, where $H_t$ is the execution time of the gate.
This is everything for the mapping of an 1-qubit gate.
Note that the execution time and fidelity of a quantum gate is provided from the building block layer.

The present work deals with a CNOT gate only for a 2-qubit gate.
Even though there is a case when a SWAP gate is required, it is possible to implement a SWAP gate with three CNOT gates.
We deal with a CNOT gate as a local gate acting on two qubits located in nearest neighbor.
Suppose that a CNOT gate is applied to qubits $q_a$ and $q_b$.
Then, to execute the gate, both qubits have to be in temporary and spatially ready status.
If the are apart, we have to move both qubits to be in neighbor via SWAP operations.
If one qubit is being manipulated by other operation, we have to delay the CNOT operation until both qubits are in idle status.
Then, the CNOT operation definitely begins at $\max \{t(q_a), t(q_b) \}$, and finishes at time $\max \{t(q_a), t(q_b) \} + CNOT_t$.
Note that $\max \{t(q_a), t(q_b\}$ is the time both qubits become idle status, and $CNOT_t$ is the execution time of a CNOT gate.

The third type quantum instruction, a module, seems like a multi-qubit composite quantum operation.
Therefore, on the surface, it seems that the mapping of a module is very similar with the mapping of a 2-qubit CNOT gate.
For the mapping of a module, argument qubits for a module should be temporally and spatially ready.
The critical difference from the case of a CNOT gate is that a distinguished physical space\footnote{The physical space for a module is the computing region we described before.} is allocated for a module.
Therefore, to perform the mapping of a module, we have to consider qubit movements between a present module and a target module.

\begin{figure}[t]
\centering
\epsfig{file=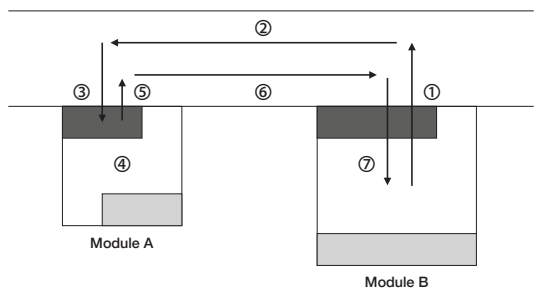, scale=0.3}
\caption{
An example of the module operation that consists of seven steps: 1. (forward) move qubits to the bus, 2. (forward) move to the target module, 3. (forward) move to the parameter qubit cells (dark grey cells), 4. module operations, 5. (backward) move qubits to the bus, 6. (backward) move to the original module, and 7. (backward) move to the original qubit positions.
}
\label{fig:parameter_passing}
\end{figure}

Suppose that a module $A$ is being mapped now, and we have to treat a quantum instruction ``$M$ ($q_a$, $q_b$, $q_c$)" that calls a module $M$ with argument qubits $q_a$, $q_b$ and $q_c$. 
Then, we have to move the argument qubits to the designated area of the module $M$. 
The qubit movements are achieved by SWAP operations through a communication bus.
We call this movement a \textit{forward} qubit passing.
After the forward qubit passing, the qubits will be placed at the parameter qubit section of the module $M$.
Please see FIG.~\ref{fig:example_module_layout} (b).
Quantum instructions of the module $M$ will then be executed.
If it is faced with a quantum instruction calling other module, then some qubits in the module $M$ will be passed to the designated space of the newly called module and manipulated there by following the quantum instructions of the module.
After executing all quantum instructions of the module $M$, the passed qubits have to be back to the original module $A$. 
We call this returning movement a \textit{backward} qubit passing.
FIG.~\ref{fig:parameter_passing} shows the module operation including the forward and backward qubit passings.


We perform the mapping for all modules sequentially as they appear in a quantum assembly code.
For that, we have to keep two lookup tables, a \textit{global} lookup table and a \textit{local} lookup table.
For each module, we first initialize a local lookup table for all qubits, and update the manipulation time of each qubit as we process each quantum instruction.
After processing all instruction of a module, we determine the execution time of the module by picking up the maximum time among all qubits.
The performance of a module is recorded in the global lookup table.
During the mapping of a module, if a module which was already mapped is called then we can refer the performance information of the called module from the global lookup table.

After mapping all modules, we can determine the execution time of a quantum algorithm as the maximum time among the qubits in the \textit{main} module. 
For example, in FIG.~\ref{fig:cat_qasm} (a), the execution time of the algorithm is $PrepZ_t$ + $FP_{main\rightarrow MakeCAT}$ + $BP_{main\leftarrow MakeCAT}$ + $MakeCAT_t$, where $MakeCAT_t=H_t + 4CNOT_t$.
Note that $FP_{main\rightarrow MakeCAT}$ ($BP_{main\leftarrow MakeCAT}$) is the time for the forward (backward) qubit passing.
The execution time of the qubit passing depends on the distance between modules.

So far, we have described a system mapping algorithm for a hierarchically structured quantum assembly code.
By the way, the presented algorithm can be applied to a non-structured quantum assembly code.
In such code, there are two types of quantum instructions: 1- and 2-qubit gate.
Regardless of the type of a quantum assembly code, as mentioned before the heart of the system mapping is 1) \textit{set up} a quantum computer architecture and 2) \textit{recast} quantum algorithm for the architecture.
To be compatible with the quantum assembly code, a simple qubit array such as a regular 2-dimensional lattice may be enough.
The proposed framework supports a system mapping on an arbitrary qubit layout as shown in FIG.~\ref{fig:mapping_simple_example} (a).

\subsection{Building Block Layer}\label{subsec:building_block}


We apply fault-tolerant quantum computing protocols based on $[[7,1,3]]$ Steane code~\cite{Steane:2006wn} and surface code~\cite{Fowler:2009ep,Fowler:2012fi}.
Both codes have well-studied logical gate protocols.
The concatenation level for Steane code and the code distance for a surface code are completely determined by a given quantum algorithm~\cite{Jones:2012kc,Suchara:2013tg}.
In this work, we set both figures by using KQ formalism~\cite{Steane:2003gp}.

\subsubsection{Steane code}\label{sec:steane_code}

$[[7, 1, 3]]$ Steane code encodes logical quantum information in a qubit into seven physical qubits, and protects it from an arbitrary 1-qubit quantum noise.
Since the transversal implementations for a logical Hadamard and a logical CNOT gate are supported, many studies on the fault-tolerant quantum computing based on the Steane code have been conducted.
In Ref.~\cite{Svore:2007tb}, an optimal design of a logical qubit for Steane code under the 2-dimensional nearest neighbor qubit interaction was proposed.
They achieved the threshold $O(10^{-5})$ with 48 physical qubits and modified quantum error correction.

In this work, we have redesigned a logical qubit with 30 physical qubits.
Seven among them are used for holding data, and the others are temporarily used for logical operations and error correction.
In particular, we applied the Shor quantum error correction~\cite{Shor:1996wi} that exploits $4$-qubit Shor state for the syndrome measurement.
For that, we prepare and verify the Shor state~\cite{Weinstein:2012jr}. 
We implemented the preparation of a logical state by following Ref.~\cite{Goto:2016jo}.
Most of logical gates are implemented as transversal gates, and the non-Clifford $T$ gate is implemented by exploiting a magic state.
We generate magic states by employing a $7$-qubit Shor state without magic state distillation~\cite{Weinstein:2015ce}.

Accuracy threshold theorem~\cite{Knill:1996tm,Aharonov:2008jn} says that if we have a quantum device of physical error rate below a code threshold, it is possible to achieve an arbitrarily reliable quantum computing. 
But, for a very large-sized quantum algorithm, encoding only once may not be enough.
Fortunately, by encoding a qubit recursively~\cite{Knill:1996ty}, we can lower the effective error rate to the degree where a reliable quantum computing is possible.

Given a quantum algorithm, we can calculate $KQ$ and determine the maximum tolerable error rate $P_{max}$ as $1/KQ$.
We then determine the concatenation level $l$ by the following inequalities satisfies
\begin{equation}
P_{max} \geq \frac{{(c_{op}p^2)}^{2^{l}}}{c_{op}},
\end{equation}
where $op$ is quantum error correction and logical operations, and $c_{op}$ is the constant factor of a specific logical operation $op$.
We obtained the constant values of each logical operation from $KQ$ of a quantum circuit for the operation.
For example, $c_{QEC}$ corresponds to $KQ$ of the QEC quantum circuit. 
In this work, we have not optimized the arrange of qubits (see Table~\ref{tab:logical_qubit_steane}), and therefore the quantum error correction and a logical operation work sub-optimally, and therefore the threshold is lower than the optimal value\footnote{Please note that the objective of our work is not to increase a code threshold, but to configure a quantum computing and analyze its performance accurately.}.

Suppose that a concatenation level for a quantum computing is determined as $l$.
The implementation of a logical $T$ gate in the level $l$ consists of only Clifford operations in a lower level $k<l$.
Then, in the level $k$, the implementation of a logical $T$ gate is not necessary and therefore the qubits to implement a magic state, the 7-qubit Shor state, are not strongly required.
Therefore only 23 qubits are required to implement a logical qubit in the lower level $k$.
But, to form a rectangular shape of a logical qubit, we require 25 qubits ($5\times 5$ layout) for a lower level qubit in the level $k$.
In Table~\ref{tab:logical_qubit_steane}, the qubits denoted by $7Sh [i]$ is not required in the lower level qubit $k$.
On the other hand, the qubits $V_{7Sh} [i]$, the main role of which is to verify the $7$-qubit Shor state, are used for the other purpose, logical measurement.

\begin{table}[t]
\caption{
The arrangement of qubits to implement a logical qubit in the concatenation level $l$.
The component qubits are in the concatenation level $l-1$.
The qubit denoted by $D[i]$ indicates a $i$-th data qubit.
The qubits $4Sh [i]$ and $7Sh [j]$ are for $4$- and $7$-qubit Shor states for syndrome measurement and a logical $T$ gate, and $V_{4Sh}$ and $V_{7Sh}$ are used to verify the $4$- and $7$-qubit Shor states respectively.
The qubit $M [i]$ is also used to implement a logical $T$ gate.
\newline
}
\small
\centering
\begin{tabular}{|c|c|c|c|c|} \hline
$V_{4Sh} [1]$ & $V_{4Sh} [2]$ & $D [1]$ & $D [2]$ & $D [3]$ \\ \hline
$4Sh [1]$ & $4Sh [2]$ & $D [4]$ & $D [5]$ & $D [6]$ \\ \hline
$4Sh [4]$ & $4Sh [3]$ & $D [7]$ & $V_{7Sh} [1]$ & $V_{7Sh} [2]$ \\ \hline
$M [1]$ & $M [2]$ & $M [3]$ & $M [4]$ & $M [5]$ \\ \hline
$M [6]$ & $M [7]$ & $V_{7Sh} [3]$ & $7Sh [1]$ & $7Sh [2]$ \\ \hline
$7Sh [3]$ & $7Sh [4]$ & $7Sh [5]$ & $7Sh [6]$ & $7Sh [7]$ \\ \hline
\end{tabular}
\label{tab:logical_qubit_steane}
\end{table}

\subsubsection{Surface code}\label{sec:surface_code}

2D surface code based fault-tolerant quantum computing is recognized as the most promising fault-tolerant quantum computing scheme due to physically less challenging requirements.
The code has a high threshold around $O(10^{-3})$~\cite{Wang:2010ts,Raussendorf:2007dr,Fowler:2009ep}, and its structure is well suited to nearest neighbor interacted qubits arranged on the 2-dimensional lattice.
In this work, we have implemented double defect based logical qubits and logical gates described in Refs.~\cite{Fowler:2009ep,Fowler:2012fi,Suchara:2013tg}.
The detailed protocols are beyond the scope of the present work, and we will describe performance parameters only.

We use the $KQ$ formalism to determine a code distance $d$~\cite{Jones:2012kc,Suchara:2013tg}.
The objective error rate of a quantum computing is determined by $P_{fail} \approx KQ\epsilon_L$, where $\epsilon_L$ is a logical error rate.
The code distance $d$ is determined as
\begin{equation}\label{eq:surface_code_distance}
d \approx \frac{2 \bigl(\log {\epsilon_L} - \log {C_1}\bigr)}{\log{C_{2}} + \log{\frac{\epsilon_p}{\epsilon_{th}}}} - 1,
\end{equation}
where $\epsilon_p$ and $\epsilon_{th}$ are physical error rate and the threshold of the surface code respectively.
$C_1$ and $C_2$ are code parameters, and we use the specific figures, $C_{1}\approx 0.13$, $C_{2}\approx 0.61$, from Ref.~\cite{Jones:2012kc}.
We apply the code threshold $\epsilon_{th}=0.009$.

Now it is possible to determine the execution time of surface code logical gates.
Above all, we need to stress that the surface code error correction has to iterate $d$ rounds of a syndrome measurement. 
We assume that logical Pauli operators are performed in classical control software by updating logical Pauli frame~\cite{Fowler:2012fi}.
A logical CNOT gate between the same type ($X$-cut or $Z$-cut) logical qubits consists of three CNOT gates between different type logical qubits.
For that, we have to prepare a pair of different type logical ancilla qubits~\cite{Fowler:2012fi}. 
A logical Hadamard gate protocol consists of cutting and reconnecting a target logical qubit from/to a whole qubit array and performing transversal physical Hadamard and SWAP gates~\cite{Fowler:2012uz,Fowler:2012fi}.
The Hadamard gate makes the role of syndrome qubits interchanged, and the syndrome qubit reverts to the original position (role) via the SWAP operation.

We now turn the attention to the non-transversal gates $S$ and $T$.
A logical $S$ gate is deterministically implemented by using a high fidelity magic state $|Y_L\rangle = \frac{1}{\sqrt{2}} \bigl(|0_L\rangle + i |1_L\rangle \bigr)$~\cite{Fowler:2012fi,Jones:2012kc}.
Since the magic state is not destroyed during the gate protocol, it is possible to reuse it if plenty of high fidelity magic states are prepared at the beginning.
Therefore before running a quantum algorithm, we prepare a number of $|Y_L\rangle$ states.
We include the duration of preparing the states in the execution time of a quantum computing.
How many $|Y_L\rangle$ states should be prepared will be discussed later.

A logical $T$ gate is implemented by consuming a high fidelity magic state $|A_L\rangle = \frac{1}{\sqrt{2}} \bigl(|0_L\rangle + \exp^{i\pi/4} |1_L\rangle \bigr)$~\cite{Fowler:2012fi}.
A magic state has to be prepared for every $T$ gate.
We assumed that a magic state is prepared and supplied in offline. 
In other words, the preparation of a high fidelity $|A_L\rangle$ is not included in the quantum computing time.
On the other hand, the logical $T$ gate operation is probabilistically achieved up to the logical $S$ gate correction.
Therefore to implement a logical $T$ gate, a $|Y_L\rangle$ state is probabilistically required.

We now discuss the quantity of the required $|Y_L\rangle$ states.
It depends on the quantity of the states maximally required at one time.
Since a logical $T$ gate probably requires a $|Y_L\rangle$, we have to prepare $|Y_L\rangle$ as much as $\max \{ parallel T, parallel S\}$ where $parallel T$ ($parallel S$) is the number of $T$ ($S$) gates executed in parallel.
Note that the quantities of $parallel T$ and $parallel S$ can be found from the system mapping process.

The preparation of a high fidelity magic state takes two steps, \textit{state injection} and \textit{state distillation}~\cite{Fowler:2012fi,Fowler:2013ht}.
The state injection in the surface code quantum computing injects an arbitrary logical state into the distance 1 logical qubit called a short qubit and makes the logical qubit larger~\cite{Fowler:2012fi}.
Enlarging a double defect logical qubit consists of multi-cell qubit movements and measurement on data qubits.
The state distillation protocol takes $m$ noisy states and generates $k$ less noisy states, where $m>k$.
By performing multiple rounds of the distillation, the magic spreaded over many states are concentrated on only a few states and therefore we can obtain high fidelity magic states.
In this work, we deal with the magic state distillation protocols described in Refs~\cite{Fowler:2009ep,Fowler:2012fi}.
The required iteration of the protocol is completely determined by the objective fidelity and a physical error rate~\cite{Suchara:2013tg}.
We set the objective error rate of the magic states as $10^{-12}$ to achieve high fidelity for the configured quantum computing\footnote{$1/ \# \textrm{T gates}$}, and empirically the 2-round distillation achieved the objective error rate in the physical error rate $10^{-3}\sim 10^{-5}$.

We determine the capacity of a magic state factory that prepares and supplies $|A_L\rangle$ states.
The capacity depends on a quantum algorithm and the durations of a state distillation and a logical $T$ gate.
Suppose that a logical $T$ gate is applied consecutively to a qubit.
Then, a magic state factory has to supply high fidelity magic states continuously.
If a magic state factory generates only one magic state at a time, there may happen a latency for the supply of the magic states if the magic state distillation takes more time than the duration of a logical $T$ gate.
Therefore, the magic state factory has the capacity to prepare at least $\max \{parallel T\} \times time(MSD)/ time(T)$ states at a time where $time(MSD)$ and $time(T)$ are the durations of the magic state distillation and the logical $T$ gate protocol.
Empirically, the $time(MSD)/time(T)$ is approximately 20 in our estimation.


To conclude, the required physical qubits for $|A_L\rangle$ and $|Y_L\rangle$ are respectively
\begin{equation}
\max \{parallel T\} \cdot \frac{time(MSD)}{time(T)} \cdot \bigl(15 \cdot Q_L)^{r-1} \cdot (16\cdot Q_L),
\end{equation}
and 
\begin{equation}
\max \{parallel T, parallel S \} \cdot \bigl(7\cdot Q_L\bigr)^{r-1}\cdot (8\cdot Q_L),
\end{equation}
where $Q_L$ is the number of physical qubits to implement a logical qubit and $r$ is the required iterations.
The last distillation round requires one more logical qubit from the Bell state~\cite{Fowler:2012fi}.
Above this, the ancilla qubits to perform CNOT gates during the distillation protocol also should be included.

\section{Performance Metric}\label{sec:performance_metric}
We describe how to evaluate the quantum computing metrics, execution time, fidelity and the quantity of qubits.

\subsection{Execution Time}\label{subsec:execution_time}

We examine the quantum computing time in two steps.
In the system mapping, we obtain the single round execution time $T_{one}$ of a quantum algorithm.
At the same time, the fidelity $F_{alg}$ of a quantum computing can be determined.
Note that how to calculate the fidelity of a quantum computing is described in the following section. 
Since $T_{one}$ is the time for running a quantum algorithm once, and there is no guarantee about a reliable quantum computing.
Noisy components may make a quantum computing broken.
To overcome the problem, we calculate the average execution time $T_{avg}$ by reflecting the number of the required iterations to achieve the fidelity 1 as
\begin{equation}
T_{avg} = T_{one}/F_{alg}.
\end{equation}
We believe this averaged time shows the time required for getting a reliable answer from a quantum computing\footnote{This does not indicate that the output from a quantum computing is an exact solution. We do not consider the probabilistic nature of a quantum algorithm.}.

\subsection{Fidelity}\label{subsec:fidelity}
The fidelity of a quantum computing can be calculated based on the fidelity of logical quantum gates as follows~\cite{Suchara:2013tg},
\begin{equation}
F_{alg} = \prod_{g} {F_g}^{N_g},
\end{equation}
where $g$ is a quantum gate utilized in the algorithm.
$F_g$ is the fidelity of the gate $g$, and $N_g$ is the total count of the gate in the algorithm.
The value $N_g$ can be found from the system mapping and $F_g$ is determined in the building block layer. 
By the way, this fidelity calculation is only applicable to Steane code based quantum computing.
As shown in Section~\ref{sec:surface_code}, the final fidelity of a surface code based quantum computing is given by $F_{alg} = 1- KQ\cdot \epsilon_L$.

\subsection{The Number of Physical Qubit}\label{subsec:qubits}

We examine the quantity of physical qubits required to run a quantum algorithm.
Since the quantity of the required qubits differs according to a fault-tolerant quantum computing scheme, we first identify the common factor, the qubits in a quantum algorithm, and then go inside specific cases later.

The proposed hierarchical quantum computer structure consists of multiple modules and a communication bus connecting all modules.
In the quantum assembly code, we can find the quantity of logical (or physical) qubits for a module.
\begin{equation}
Q_{comp} = \sum_{M} \bigl(Q^{M}_{local} + Q^{M}_{param} \bigr), 
\end{equation}
where $Q^{M}_{local}$ ($Q^{M}_{param}$) is the number of local (parameter) qubits of a (computing) module $M$.

\subsubsection{Steane Code Quantum Computing}
We consider the Steane code quantum computing.
The structure of a communication bus depends on the chosen global layout over all modules.
On the 1D global layout, the number of the qubits can be simply calculated as $Q_{comm} = bandwidth \times length$, where $length$ is obtained as 
\begin{equation}
length = \sum_{M} M_{width},
\end{equation}
where $M_{width}$ is the width of a module, which is 1 for 1D local layout in common and $\lfloor \sqrt{Q^{M} } \rfloor$ for 2D local layout.
Note that $Q^{M} = Q^{M}_{local} + Q^{M}_{param}$.

On the other hand, on the 2D global layout, the number of qubits can be calculated as follows.
Let us suppose that the number of modules is $|M|$.
Then, $\lfloor \sqrt{|M|} \rfloor \times \lfloor \sqrt{|M|} \rfloor$-sized 2D layout is necessary.
To keep the shape of a module on the 2D layout, all modules have the same size cells $n\times n$, where $n = \lfloor \sqrt{\max \{Q^{M}}\} \rfloor$.
Then, the required logical qubits for the communication bus is 
\begin{equation}
Q_{comm} = 2\cdot bandwidth\cdot n \cdot A\cdot B + \bigl(n \cdot A\bigr)^2
\end{equation}
where $A = \lfloor \sqrt{|M|} \rfloor - 1$ and $B = \lfloor \sqrt{|M|} \rfloor$.
In this work, we determine the bandwidth of a bus as the maximum number of parameter qubits, $bandwidth = \max_{M} \{Q^{M}_{param}\}$.

So far, we have identified the number of logical qubits required for a quantum computing.
As we mentioned in Section~\ref{sec:steane_code}, a logical qubit in the concatenation levels $k=1\sim l-1$ is composed of 25 qubits and the qubit in the level $l$ consists of $30$ qubits.
According to a physical error rate, we need to apply recursive encoding.
Therefore, the number of total physical qubit in the Steane code quantum computing is 
\begin{equation}
Q_{steane} = 25^{r-1} \cdot 30 \cdot Q_{comp} + 25^r \cdot Q_{comm} 
\end{equation}
where $r$ is the determined concatenation level.

\subsubsection{Surface Code Quantum Computing}
\begin{figure}[t]
\centering
\epsfig{file=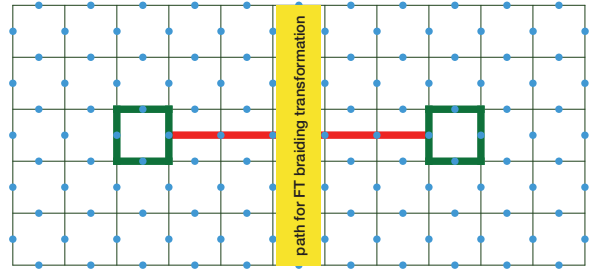, scale=0.28, angle=90}
\caption{
	A double $Z$-cut qubit of a code distance 3.
	The blue dots indicate data qubits.
	One of the green chains indicates a logical $Z$ operator, and the red chain indicates a logical $X$ operator.
	Through the yellow line, it is possible to perform a fault-tolerant braiding operation from other $X$-cut qubit to this $Z$-	cut qubit. 
	Each defect has to be away from boundary as much as 3 data qubits, and both defects have to be separated 6 data qubits.
	126 data qubits and 125 syndrome qubits are required to implement a distance-3 logical qubit.
}
\label{fig:double_defect_distance_3}
\end{figure}

We implement double defect based logical qubits. 
For a double defect logical qubit with code distance $d$, each defect has to be apart from a boundary as much as $d$ data qubits, and double defects also should be separated as much as $d$ data qubits.
On the other hand, to perform a braiding operation in a fault tolerant manner, the space between double defects has to be at least $2d+\lfloor d/4 \rfloor$ rather than only $d$.
Therefore, to implement a double defect logical qubit of code distance $d$, $(2A+1)(2B+1)$ physical qubits are required where $A=\bigl(2d-2 + \lfloor d/4 \rfloor \bigr)$ and $B=\bigl(4d-4 + 3\lfloor d/4 \rfloor  \bigr)$.
FIG.~\ref{fig:double_defect_distance_3} shows a double defect logical qubit of code distance 3.
A total of 253 physical qubits, 126 data qubits and 125 syndrome qubits, are required.

Two neighboring logical qubits also have to be separated as much as $\lfloor d/4 \rfloor$ data qubits to keep the code distance between both during the braid transformation.
In this regards, if $N$ logical qubits are arranged on the 2-dimensional layout of $n_h\times n_w$, we need 
\begin{equation}
\Bigl(2\bigl(n_w A + (n_w-1)\lfloor d/4 \rfloor\Bigr)+1\Bigr) \Bigl(2\bigl(n_h B + (n_h -1)\lfloor d/4\rfloor \bigr)+1\Bigr)
\end{equation}
qubits are necessary, where $A$ and $B$ are what we mentioned above.


We have to consider ancilla qubits required for a CNOT gate. 
As mentioned before, the CNOT gate between the same type logical qubits consists of three CNOT gates between different logical qubits.
For that two ancilla qubits, $X$-cut qubit $|g_L\rangle$ and $Z$-cut qubit $|+_L\rangle$ are required.
We allocate a pair of both ancilla qubits to every module where a CNOT gate is performed.
In that case, the number of logical qubits for a module is the sum over parameter qubits, local qubits and two ancilla qubits.

\begin{figure}[t]
\centering
\epsfig{file=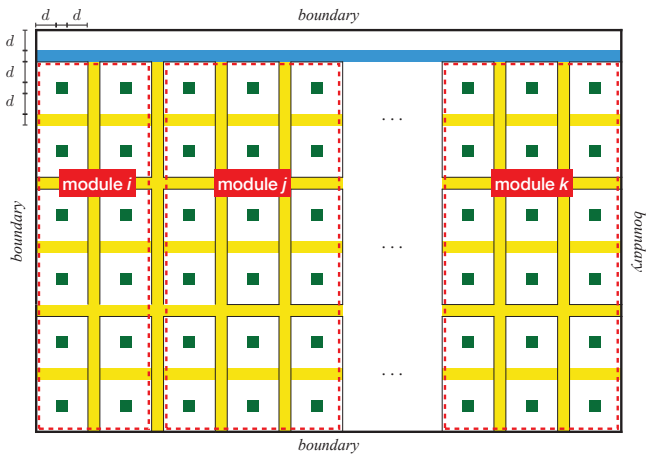, scale=0.3}
\caption{
	A quantum computer structure based on the surface code quantum computing.
	Dark green dots indicate defects and yellow cells are used as a path for the braiding transformations.
	Blue cells can be used for the forward/backward qubit passings over distance modules.
	An enclosed section by dotted red line is a computing region, a module.
	A defect has to be away from the boundary of a logical qubit or a braiding path as much as $d$ data qubits, and two logical qubits are mutually separated as much as $\lfloor d/4 \rfloor$ data qubits.
}
\label{fig:surface_code_architecture}
\end{figure}

In case of a surface code quantum computing, the communication bus may be also necessary to efficiently perform forward/backward qubit passings\footnote{It is possible to perform a fault-tolerant braiding between distant logical qubits in different modules. On considering that, the qubit passings may not be required. By the way, a braiding between distant logical qubits requires so many physical measurements. Which makes a quantum computing unreliable. In this regards, we believe that moving logical qubits to nearby location (a target module) and performing a braiding between close qubits is more reliable. In this regards, we also perform qubit passing in the surface code quantum computing.}.
By the way, unlike the Steane code quantum computing that performs a sequence of SWAP operations as much as the passing distance, the qubit movement in the surface code quantum computing is much efficient.
It can be achieved by only performing multi-cell qubit movements~\cite{Suchara:2013tg,Fowler:2012fi}.
Therefore, we assumed that the surface code quantum computing performs the qubit passing in sequentially on the bus of the narrow bandwidth.
We set the bandwidth of the bus as $\lfloor d/4 \rfloor$, and additionally the movement path should be away from a boundary as much as $d$ data qubits.
FIG.~\ref{fig:surface_code_architecture} shows the quantum computer architecture based on a surface code and a structured quantum assembly code, where all the modules are arranged on the 1-dimensional layout by keeping the space as much as $\lfloor d/4 \rfloor$ data qubits between both modules.

We conclude this section by repeating the physical qubits for the magic state factory. 
The required physical qubits for $|Y_L\rangle$ are $\max \{parallel T, parallel S \}\times (7Q_L)^{r-1}\times (8Q_L)$ and for $|A_L\rangle$ are $\max \{parallel T\} \times (15Q_L)^{r-1}\times (16Q_L)\times time(MSD)/time(T)$.
Note that $Q_L$ is the physical qubits to implement a logical double defect qubit, and $r$ is the iteration of the magic state distillation.

\section{Analysis of Performance and Resource}\label{sec:performance_evaluation}

We show the performance analysis of quantum computings we configured.
For that, we set the objective fidelity of a quantum computing as 0.7.
That is, a single round quantum computing time $T_{one}$ is the time of a quantum computing that achieves a fidelity at least 0.7.
We set the error rate of physical operation for Steane code quantum computing as $10^{-9}$, but for surface code quantum computing we apply the physical error rate $10^{-3}$.
We also assume that the duration of a physical operation is 1 $\mu s$ conservatively.
This assumption may be pessimistic than other literature assuming tens $\sim$ hundreds nano seconds for a physical operation.
The Shor algorithm we test comes from the benchmark~\cite{ScaffCCScaffCC:vm,Suchara:2013tg}.

\subsection{Case of applying Compile}\label{sec:performance_compile}

We show the performance changes by applying a quantum compile, i.e., decomposition of a $R_Z(\theta)$ gate for an arbitrary angle $\theta$.
Even though such decomposition is required to implement a fault-tolerant quantum computing, in this section we perform physical quantum computing without error correction to see the effect of the compile only.
For that, the components of the other layers are completely fixed.

For the decomposition, we set the precision of the decomposition as $10^{-2}$, which means that a decomposition of $R_Z(\theta)$ gate achieves the $R_Z(\theta)$ operation with an error probability $10^{-2}$.
Consequentially, both $R_Z(\theta_1)$ and $R_Z(\theta_2)$ can be decomposed into the same sequence of $H$, $S$ and $T$ if $|\theta_1-\theta_2| \leq 0.01$.
Under such precision, a $R_Z(\theta)$ gate is usually decomposed into a sequence of $40\sim 50$ $H$, $S$ and $T$ gates.\footnote{If we set the precision degree with a smaller number, we will get a longer sequence of $H$, $S$ and $T$ gates for a $R_Z(\theta)$ gate. On the one hand, such sequence can achieve the target $R_Z(\theta)$ gate more exactly. On the other hand, the quantum computing time will be larger than the time shown in this work. Besides, practically the duration to conduct the performance analysis also increase nontrivially. In this regards, we have set $10^{-2}$ for the precision.}
Please note that the decomposition algorithm in the compile works probabilistically.


\begin{figure*}[t]
\centering
\subfigure[]{
	\epsfig{file=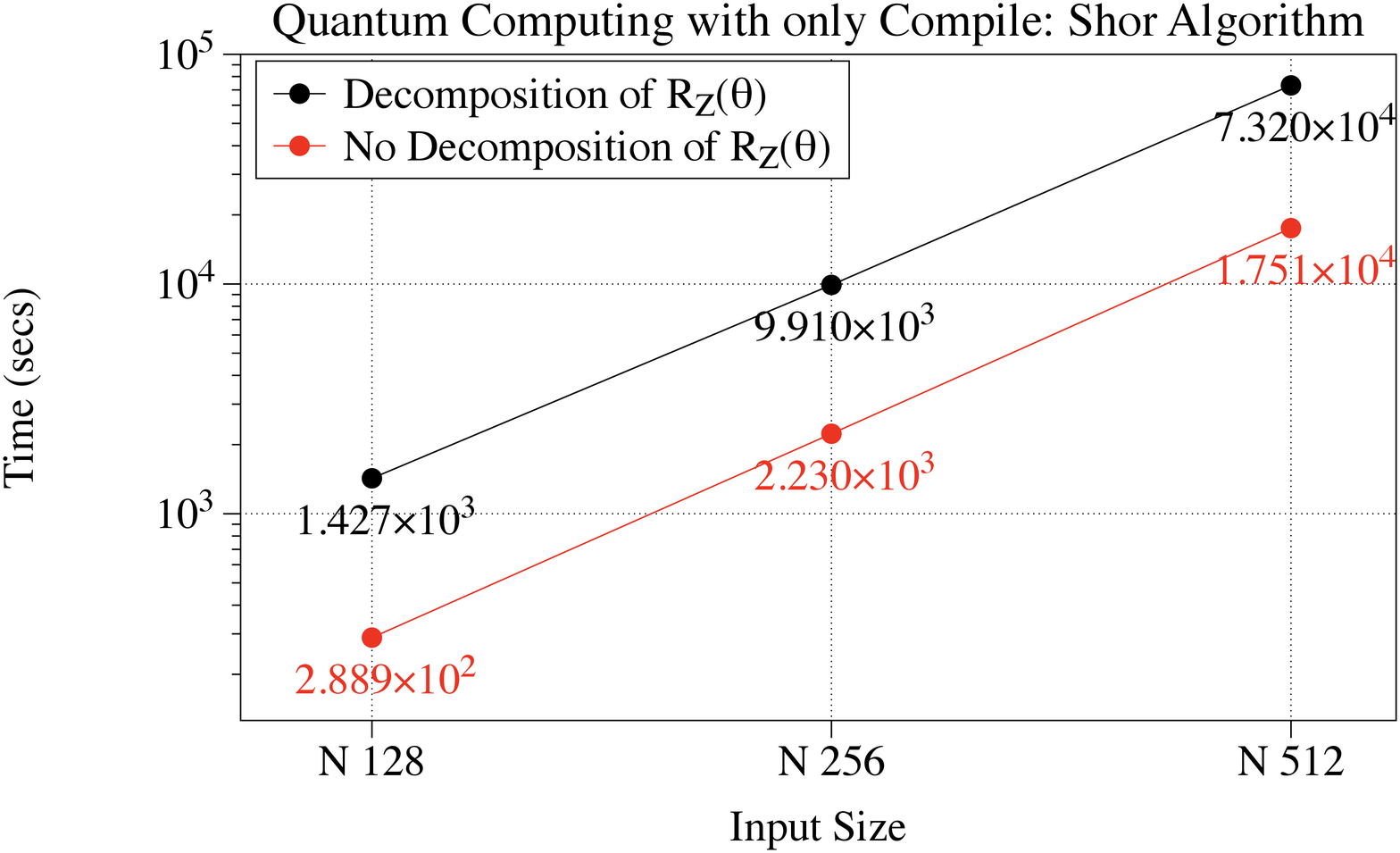, scale=0.22}
}
\subfigure[]{
	\epsfig{file=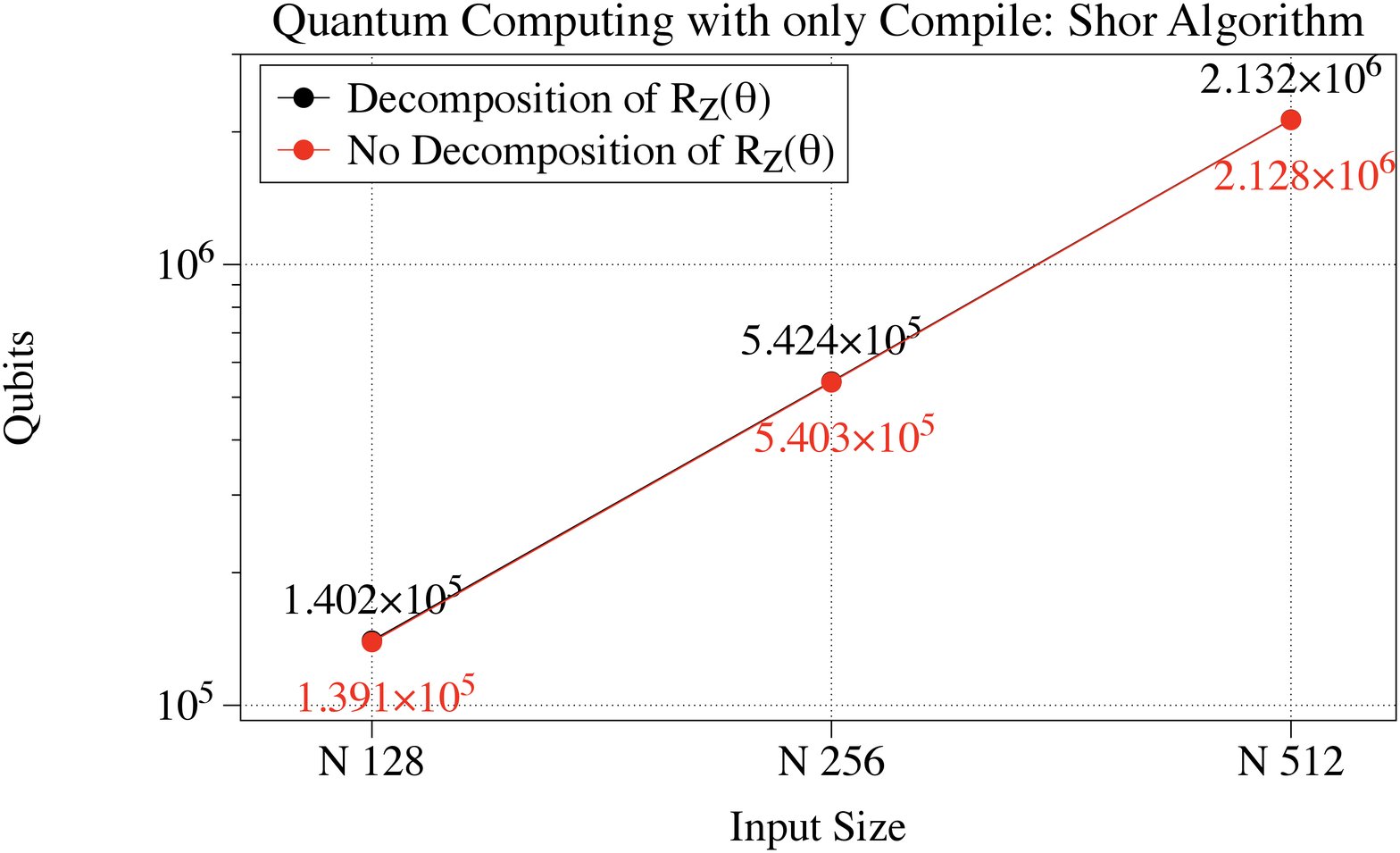, scale=0.22}
}
\caption
{
We show the quantum computing performance change by the compile effect. 
(a) Quantum computing time and (b) Qubits.
}
\label{fig:noiseless_nonlocal_gate}
\end{figure*}

FIG.~\ref{fig:noiseless_nonlocal_gate} compares the performance. 
By decomposing $R_Z(\theta)$ gate, the quantum computing time increases as much as $4\sim 5$ times, but the number of physical qubits stays equivalently.
By the way, in general $R_Z(\theta)$ gate takes more than half of all quantum gates in Shor's factoring algorithm (see Table~\ref{tab:RZ_ratio}).
On considering that $R_Z(\theta)$ gate is decomposed into a sequence of dozens of $H$, $S$ and $T$ gates as we mentioned above, readers may guess that the performance difference between both cases should be more larger than the shown in the figure.

As mentioned above, we have set the precision of the decomposition as $10^{-2}$.
Most $\theta$ in the Shor algorithm are very small ($< 0.01$),\footnote{$\theta = \pi/2^{n-1}$ with $n=1\sim N$ for $N$-bit integer factoring.} and therefore the decomposition of such rotation operation works as the identity operation. 
We show the top dominant $\theta$ used in Shor N=128 algorithm in Table~\ref{tab:dominant_theta}.
All the angles are less than $0.01$.
While we have not described all $\theta$ in the algorithm in the table, empirically $75\%$ of the angles applied in the algorithm are less than $0.01$.
In this regards, the performance degradation by decomposing $R_Z(\theta)$ gates is not so remarkable regardless of the quantity of $R_Z(\theta)$ gates in the algorithm.

\begin{table}[t]
\caption{
The proportion of $R_Z(\theta)$ gate in Shor $N=128$.
\newline
}
\small
\centering
\begin{tabular}{c|c|c|c} \hline 
Input Size & $R_Z(\theta)$ & Total Gates & Proportion\\ \hline
128 & $2.036 \times 10^{9}$ & $3.399\times 10^{9}$ & 59.90\%\\ \hline
256 & $1.630\times 10^{10}$ & $2.719\times 10^{10}$ & 59.94\%\\ \hline
512 & $1.304\times 10^{11}$ & $2.175\times 10^{11}$ & 59.95\%\\ \hline
\end{tabular}
\label{tab:RZ_ratio}
\end{table}

\begin{table}[t]
\caption{
List of top 10 dominant angles in Shor's factoring algorithm, N=128.
The $\theta$ listed in this table is less than $0.01$ and therefore $R_Z(\theta)$ works as an identity operator.
The rotational angle $\theta$ of the gate is from $\pi/2^{n-1}$ in Quantum Fourier Transform, and the exact representation of the angle is limited by a classical computer precision.
\newline
}
\small
\centering
\begin{tabular}{c|c|c} \hline 
$\theta$ & Count & Proportion\\ \hline
$0.000000\times 10^{0}$ & $6.88\times 10^{8}$ & $0.3381$ \\ \hline
$-0.000000\times 10^{0}$ & $3.44\times 10^{8}$ & $0.1691$ \\ \hline
$-5.000000\times 10^{-5}$ & $3.13\times 10^{7}$ & $0.0154$ \\ \hline
$-1.000000\times 10^{-4}$ & $3.11\times 10^{7}$ & $0.0153$ \\ \hline
$5.000000\times 10^{-5}$ & $3.10\times 10^{7}$ & $0.0152$ \\ \hline
$-2.000000\times 10^{-4}$ & $3.09\times 10^{7}$ & $0.0152$ \\ \hline
$1.000000\times 10^{-4}$ & $3.08\times 10^{7}$ & $0.0151$ \\ \hline
$-4.000000\times 10^{-4}$ & $3.08\times 10^{7}$ & $0.0151$ \\ \hline
$2.000000\times 10^{-4}$ & $3.07\times 10^{7}$ & $0.0151$ \\ \hline
$-7.500000\times 10^{-4}$ & $3.06\times 10^{6}$ & $0.0150$ \\ \hline
\end{tabular}
\label{tab:dominant_theta}
\end{table}

\subsection{Case of applying Compile and Error Correction}\label{sec:performance_qec}


We show the performance of a fault-tolerant quantum computing but without considering local qubit interaction on a quantum computer architecture.
We assume that all qubits are directly interacted with an arbitrary qubit regardless of its position, and therefore a communication bus is not required in the architecture.
This setting is to see the effect of quantum error correction.
For that, we only configured Steane code based quantum computing because surface code computing inherently takes 2-dimensional nearest neighbored qubit layout into consideration.
As mentioned above, for the fault-tolerant quantum computing, we compile the quantum algorithm by decomposing $R_Z(\theta)$ gate into $H$, $S$ and $T$ gates.


\begin{figure*}[t]
\centering
\subfigure[]{
	\epsfig{file=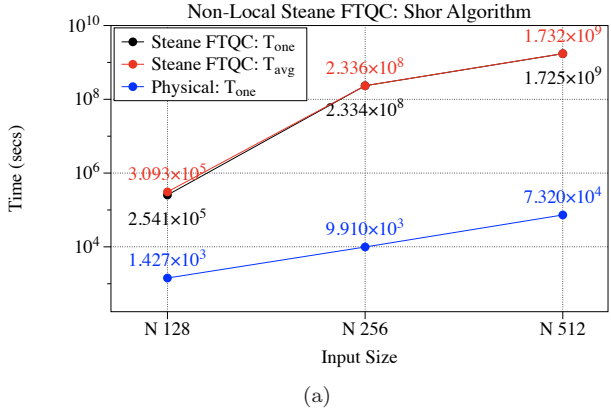, scale=0.22}
}
\subfigure[]{
	\epsfig{file=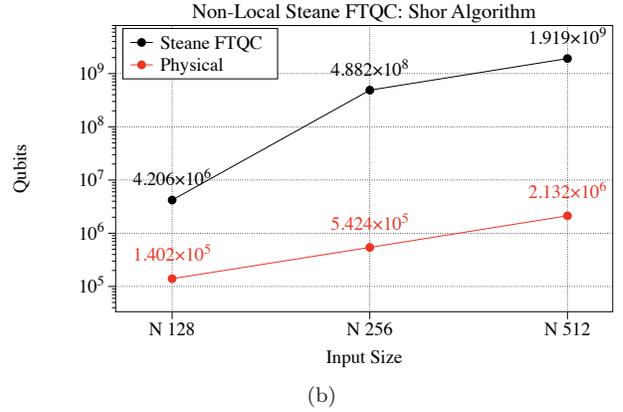, scale=0.22}
}
\caption
{
Quantum computing performance based on quantum error correction.
For the fault-tolerant operation, we have compiled a quantum algorithm by decomposing $R_Z(\theta)$ gates into a sequence of $H$, $S$ and $T$ gates.
In this evaluation, the quantum computer architecture and local qubit interaction are not completely considered.
(a) Quantum computing time and (b) Qubits.
The concatenation level for the input size 128 is 1, and 2 for the other cases.
}
\label{fig:noisy_nonlocal_gate}
\end{figure*}

FIG.~\ref{fig:noisy_nonlocal_gate} shows the quantum computing performance.
The increase of the execution time and the number of qubits is very remarkable when the input size increases from 128 to 256.
This is because the required concatenation level increases from 1 to 2 there to satisfy the objective fidelity 0.7.
But, as the concatenation level stays when the input increases from 256 to 512, the increases of a quantum computing time and the number of qubits are rather modest.

Since in this section we assume a fault-tolerant quantum computing but with non-local qubit interaction, the performance change shown in the figure is only caused by the fault-tolerant quantum computing protocol.
For example, in FIG.~\ref{fig:noisy_nonlocal_gate} (b), the numbers of qubits in the Steane code quantum computing are bigger than physical computing as much as respectively 30, 900 and 900 times.
Please recall that we have designed a logical qubit by assembling 30 physical qubits.

\subsection{Case of applying Compile, Error Correction and System Architecture}\label{sec:performance_local_qec}

In this section, we show the quantum computing performance by considering all the realistic factors we have described previously.
We apply fault-tolerant quantum computings based on certain quantum computer architectures where only local qubit interaction is permitted.

\begin{figure*}[t]
\centering
\subfigure[]{
	\epsfig{file=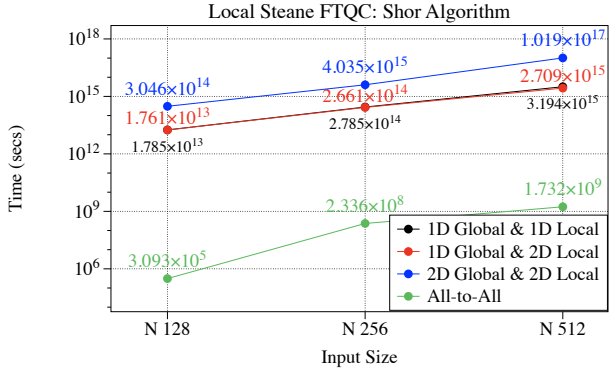, scale=0.22}
}
\subfigure[]{
	\epsfig{file=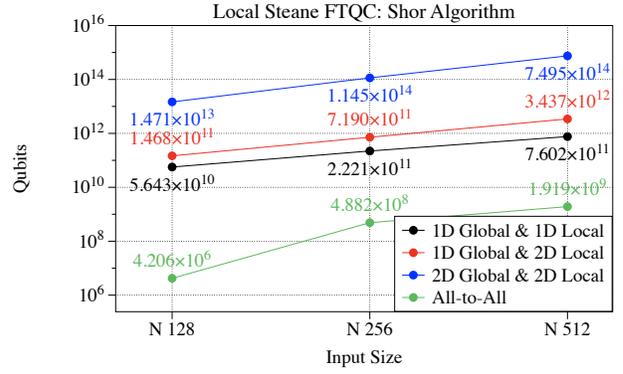, scale=0.22}
}
\caption
{
Quantum computing performance of the Steane code based local fault-tolerant quantum computing.
Too see the influence by the architectural limitation, we also add the performance when the arbitrary long qubit interaction is allowed.
(a) Quantum computing time and (b) Qubits.
All concatenation levels for the local qubit interaction cases (black, red and blue lines) are 3 in common. 
But, in case of the non-local qubit interaction (green line), the concatenation level is 1 for the input size 128 and 2 for the others.
Please see FIG.~\ref{fig:layout_example} about the quantum computer architectures.
}
\label{fig:noisy_local_gate_steane}
\end{figure*}

FIG.~\ref{fig:noisy_local_gate_steane} shows the performance analysis of the Steane code quantum computing.
We have used the quantum computer architectures of the layouts; (1D global, 1D local), (1D global, 2D local) and (2D global, 2D local).
Please see FIG.~\ref{fig:layout_example} for the quantum computer architectures.
To see the influence of local qubit interaction, we also compare the performance of the quantum computing based on non-local qubit interaction shown in the previous section.

As shown in the figure, the performance degradation by the local qubit interaction on a quantum computer architecture is highly nontrivial.
This is because many modules are spreaded over the quantum computer, and the communication (qubit passing) are performed frequently.
Table~\ref{tab:SWAP_ratio} shows the proportion of SWAP gates in the implementation of Shor algorithm.
Surprisingly, on the proposed quantum computer architecture with a nearest neighbor qubit interaction, most of quantum operations in the Steane code quantum computing are qubit movements.
We think the quantity of the qubit movements is a temporal overhead to implement a quantum algorithm on a quantum computer.
Such large overhead caused by the qubit movements can be reduced by improving a quantum computer structure, a fault-tolerant quantum computing scheme or a system mapping algorithm.

\begin{table}[t]
\caption{
The proportion of \textit{SWAP} gate in Shor's factoring algorithm.
The layout indicates a combination of Global Layout and Local Layout.
\newline
}
\small
\centering
\begin{tabular}{c|c|c|c|c} \hline 
Input Size & Layout & SWAP & Total Gates & Proportion\\ \hline
\multirow{3}{*}{128} & (1D, 1D) & $ 7.371\times 10^{11}$ & $ 7.405\times 10^{11}$ & 99.54\%\\ \cline{2-5}
 & (1D, 2D) & $1.262\times 10^{12}$ & $1.266\times 10^{12}$ & 99.73\%\\ \cline{2-5}
 & (2D, 2D) & $1.527\times 10^{13}$ & $1.527\times 10^{13}$ & 99.97\%\\ \hline
\multirow{3}{*}{256} & (1D, 1D) & $1.116\times 10^{13}$ & $1.118\times 10^{13}$ & 99.76\%\\ \cline{2-5}
 & (1D, 2D) & $1.856\times 10^{13}$ & $1.859\times 10^{13}$ & 99.85\%\\ \cline{2-5}
 & (2D, 2D) & $2.719\times 10^{14}$ & $2.720\times 10^{14}$ & 99.99\%\\ \hline
\multirow{3}{*}{512} & (1D, 1D) & $1.068\times 10^{14}$ & $1.070\times 10^{14}$ & 99.80\%\\ \cline{2-5}
 & (1D, 2D) & $1.811\times 10^{14}$ & $1.813\times 10^{14}$ & 99.88\%\\ \cline{2-5}
 & (2D, 2D) & $5.789\times 10^{15}$ & $5.789\times 10^{15}$ & 99.99\%\\ \hline
\end{tabular}
\label{tab:SWAP_ratio}
\end{table} 

\begin{figure*}[t]
\centering
\subfigure[]{
	\epsfig{file=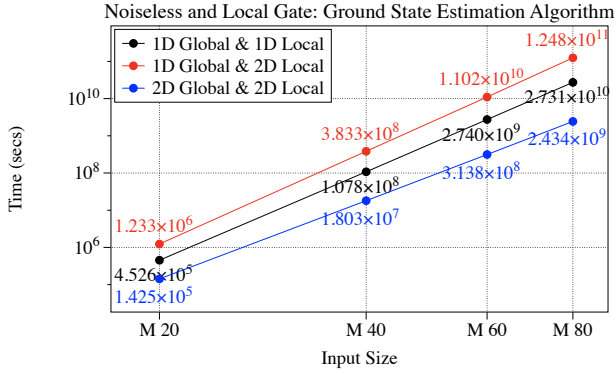, scale=0.22}
}
\subfigure[]{
	\epsfig{file=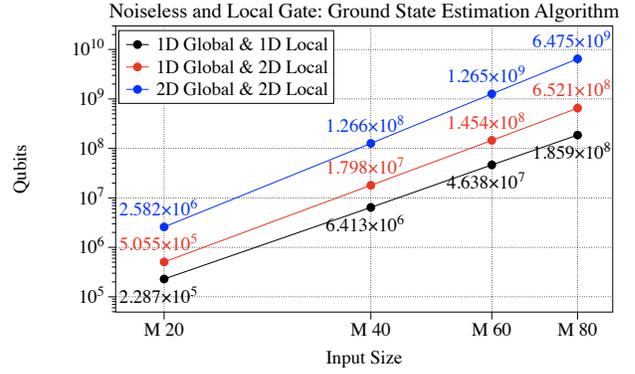, scale=0.22}
}
\caption
{
The quantum computing performance of a ground state estimation algorithms over input size $M=20, 40, 60, 80$.
Quantum gates are noiseless, and only nearest neighbor qubits are mutually interacted on the quantum computer architectures.
(a) Quantum computing time and (b) Qubits.
}
\label{fig:noiseless_local_gate_gse}
\end{figure*}

The figure shows that a quantum computer architecture of the 1D global layout provides the better performance than a quantum computer of the 2D global layout.
However, it may not always be the case.
It completely depends on the number of modules in a quantum computing program (see FIG.~\ref{fig:scaffold_code}), and the arrangement of the modules on the architecture.
In general, the 2D global layout is a better architecture in terms of a qubit communication when the number of modules is very large.
On average, the arrangement of the modules on the 2D global layout can reduce the distance between modules than the 1D global layout.
Therefore the communication cost of the qubit passing is less than the 1D global layout.
As an example, FIG.~\ref{fig:noiseless_local_gate_gse} shows that a ground state estimation algorithm~\cite{Suchara:2013tg,JavadiAbhari:2015jf,ScaffCCScaffCC:vm} works better on a quantum computer architecture with 2D global layout\footnote{In the benchmark program, the algorithm is composed of at least tens of thousands modules, but the program of Shor algorithm consists of thousands modules.}.

\begin{figure*}[t]
\centering
\subfigure[]{
	\epsfig{file=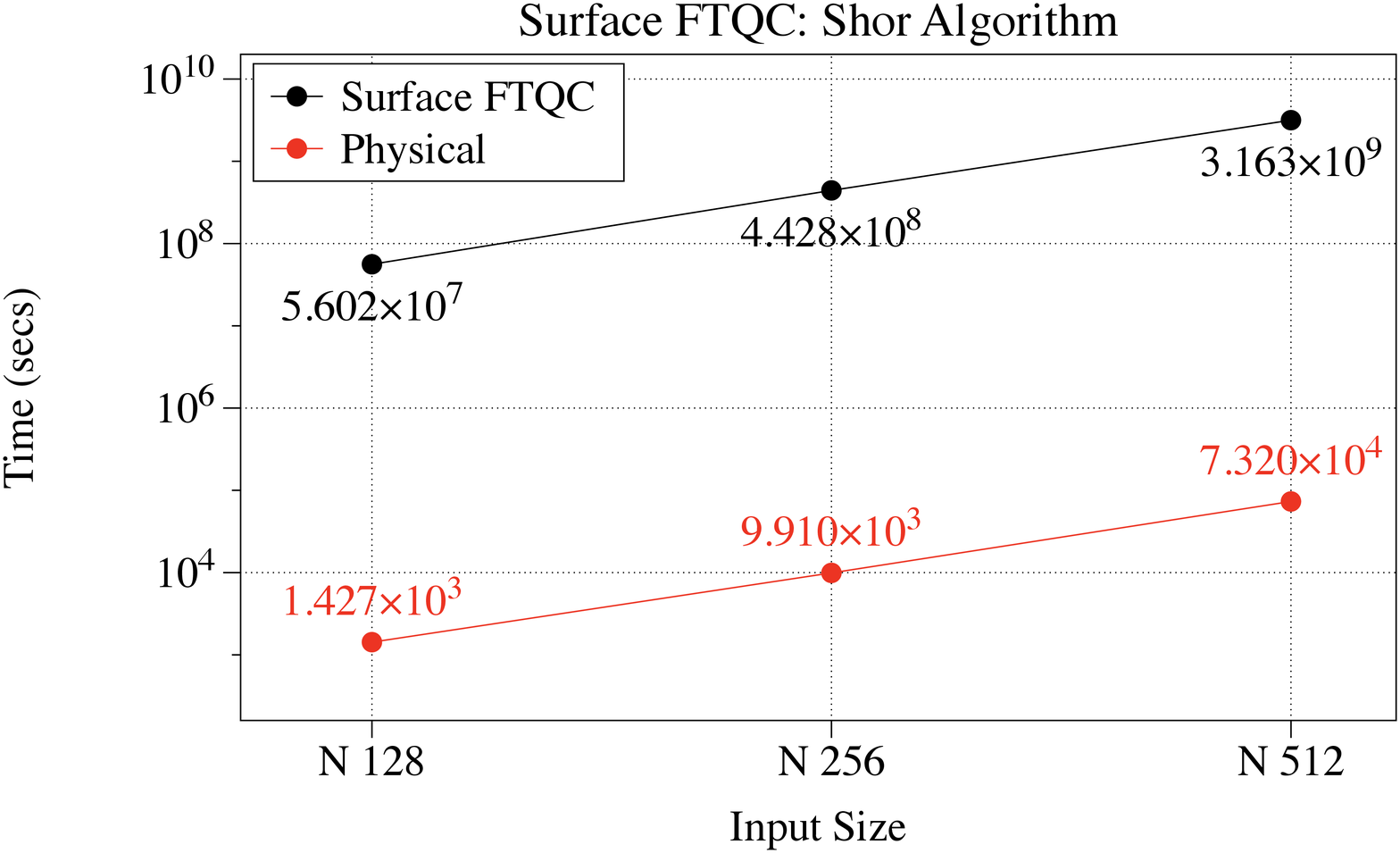, scale=0.22}
}
\subfigure[]{
	\epsfig{file=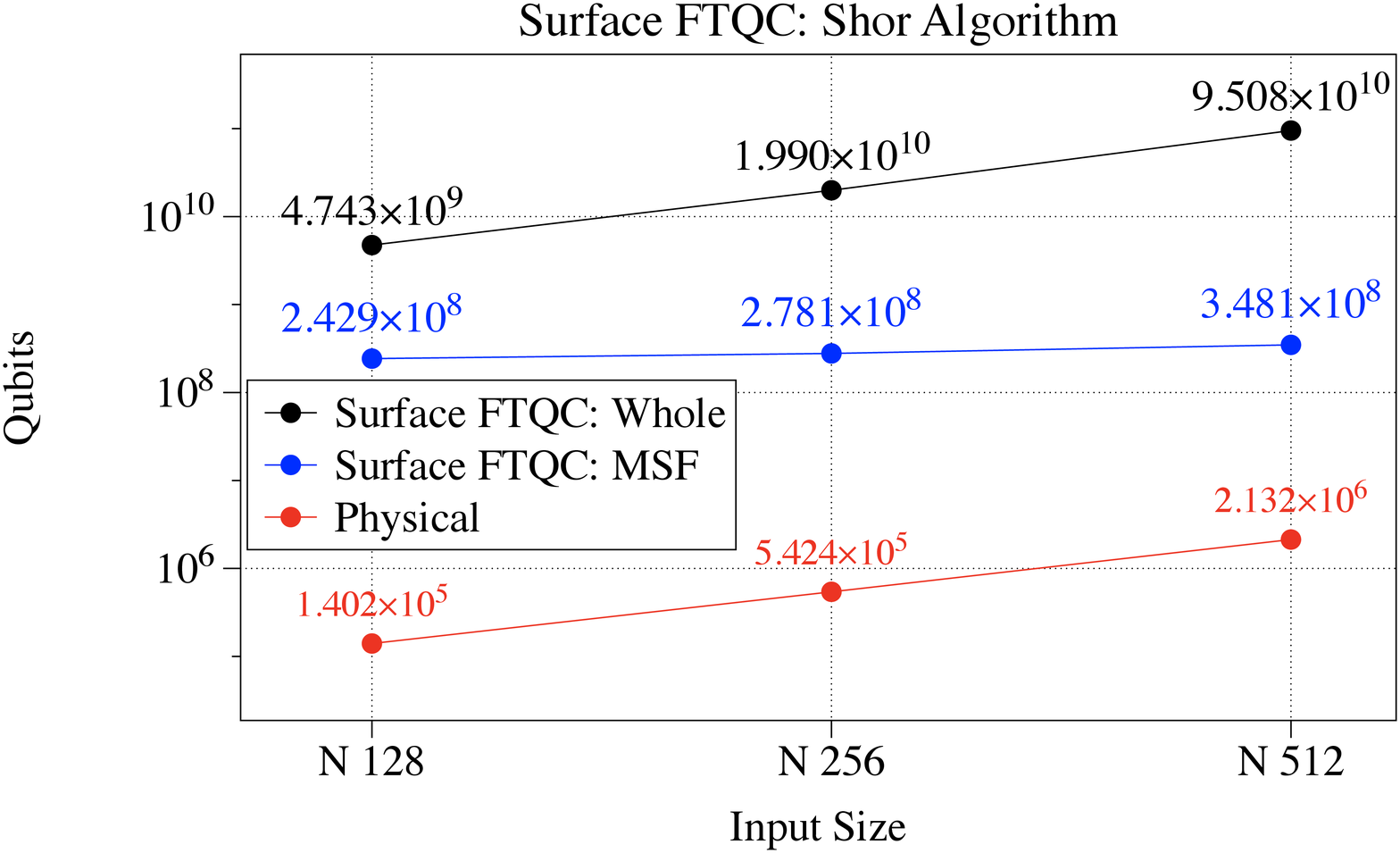, scale=0.22}
}
\caption{
The quantum computing performance based on the surface code quantum computing.
(a) Time and (b) Qubits.
The code distances are respectively 25, 27 and 30.
In (b), we also show the required physical qubits for a magic state factory.
}
\label{fig:surface_code}
\end{figure*}

FIG.~\ref{fig:surface_code} shows the quantum computing performance of the surface code quantum computing.
The quantum computer architecture for a surface code quantum computing is shown in FIG.~\ref{fig:surface_code_architecture}.
In the error rate $10^{-3}$, as the input size increases, the required code distance is raising 25, 27 and 30 to satisfy the objective fidelity.
In the figure, we also show the quantity of physical qubits to run a magic state factory that supplies $|A_L\rangle$ states during the quantum computing.
As shown in the figure, in this work, the capacity of a magic state factory stays almost the same regardless of the input size of Shor algorithm.
It increases as much as the code distance.

In Ref.~\cite{Fowler:2012fi}, the authors estimated the surface code quantum computing execution time of Shor algorithm for factoring a 2000-bit integer.
They did the estimation only by focusing on the quantity of the sequential Toffoli gates in the modular exponentiation circuit.
By following their estimation method, the quantum computing time to factorize a 512-bit integer will be only 0.45 hours ($40\times 512^3 \times 3\times100~ns$) regardless of physical error rate.
However, as shown in the figure, our estimation shows $8.78\times 10^ 5$ hours are required for the same task when the physical error rate is $10^{-3}$.
The algorithm execution time, as shown in Section~\ref{sec:accurate_gates}, is reduced as the physical error rate decreases.

We believe the reasons for such enormous gap between both estimations are as follows.
First, our analysis is based on all quantum gates included in Shor algorithm but their analysis focuses on the Toffoli gates only.
In our estimation, the proportion of transversal gates takes about 60 \% (\# transversal gates/\# total gates), in other words, their estimation ignores the execution time by such huge transversal gates and the involved error correction.
Second, we have assumed that a physical gate operates in 1 $\mu s$ conservatively while their estimation is based on 100 $ns$ measurement gates\footnote{Their implementation of a Toffoli gate completes the operation within three measurement operations.}.
Third, while they used an efficient decomposition of a Toffoli gate, we have used the de facto standard decomposition~\cite{Nielsen:2000ga}.
Fourth, they might assume that braiding operations for parallel CNOT gates can be performed in parallel without any path conflicts, but we applied a braiding operation at a time to avoid conflicts between other braiding paths.
Please note that for Shor algorithm of $N=m$, $m$ CNOT gates can be performed in parallel in ideal case.
Fifth, we have considered architecture related issues such as qubit passing over distant modules, but they do not.
Lastly, we have applied $d$ round quantum error correction where the code distance $d$ is determined based on the physical error rate, but they did not.

\begin{figure*}[t]
\centering
\subfigure[]{
	\epsfig{file=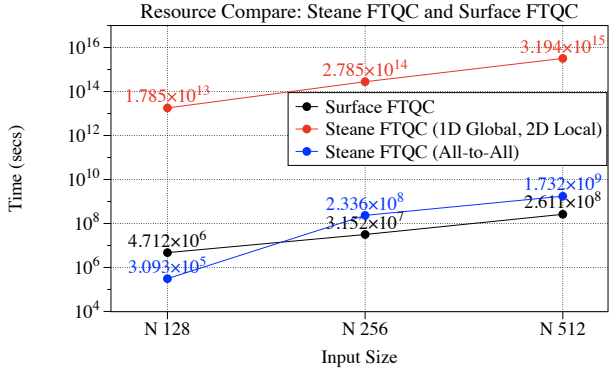, scale=0.22}
}
\subfigure[]{
	\epsfig{file=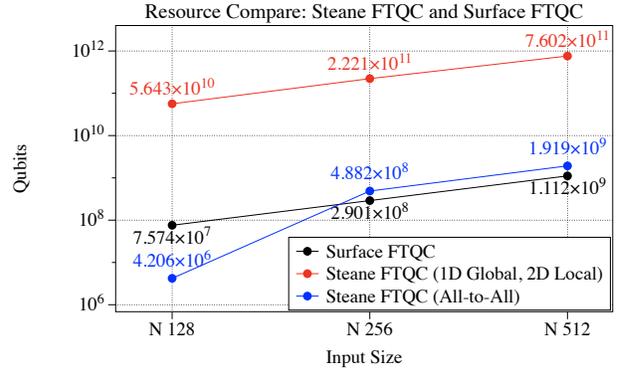, scale=0.22}
}
\caption
{
We simply compare the quantum resource in the Steane code quantum computing and the surface code quantum computing.
The physical error rate is $10^{-9}$.
The quantum computer architecture for the Steane code quantum computing is the 1D global layout and the 2D local layout because as shown in FIG.~\ref{fig:noisy_local_gate_steane} the architecture shows the best performance in this work.
We also compare the Steane code quantum computing with non-local qubit interaction because the performance of the Steane code quantum computing significantly depends on the quantum computer architecture.
}
\label{fig:steane_surface_compare}
\end{figure*}

One of the reasons why a surface code has attracted much attention is it requires relatively less overhead.
In what follows, we simply compare the Steane code and surface code in light of quantum resource without considering their theoretical foundation.
For the fair comparison, we assumed the error rate of physical device is $10^{-9}$ for both cases. 
FIG.~\ref{fig:steane_surface_compare} shows the quantum computing time and qubits to run Shor algorithm.
As we mentioned before, the performance of the Steane code quantum computing completely depends on a quantum computer architecture.
Therefore to focus on the difference in quantum resource only by a fault-tolerant quantum computing, we also compare the situation where a non-local qubit interaction is allowed.
As shown in the figure, in the small input size, the Steane code quantum computing requires less time and qubits than the surface code quantum computing.
This is because of the non-locality of multi-qubit operation.
But, as the input size increases, the surface code quantum computing shows the better performance than the Steane code quantum computing even non-local qubit interaction is allowed.


\section{Usability of the proposed framework}\label{sec:performance_improvement}

The objective of the proposed framework is to help to design and analyze a quantum computing.
In this regards, in this section, we show how to use it for analyzing previously proposed high performance quantum computing methods with realistic quantum computer. 
The first is an efficient compile (Section~\ref{sec:efficient_CRN}), and the second is an improved physical gate (Section~\ref{sec:accurate_gates}) and the last is the strategy for the fault-tolerance (Section~\ref{sec:degree_fault_tolerance}).

\subsection{Efficient Decomposition of Controlled-$R_n$}\label{sec:efficient_CRN}
Authors proposed an efficient decomposition algorithm for a controlled-$R_n$ gate~\cite{Kim:2018hs}.
By hiring an ancilla qubit, they achieved that the total number of quantum gates $\{H, S, T\}$ is reduced to from 35 (Ref.~\cite{Kliuchnikov:2013tr}) to 21.
We show how the proposed decomposition affects the execution time of Shor algorithm. 
Even though the proposed compile algorithm itself requires more qubits, by reducing the algorithm execution time and increasing the fidelity of a quantum computing simultaneously, in total less qubits are required.

\begin{figure*}[t]
\centering
\subfigure[]{
	\epsfig{file=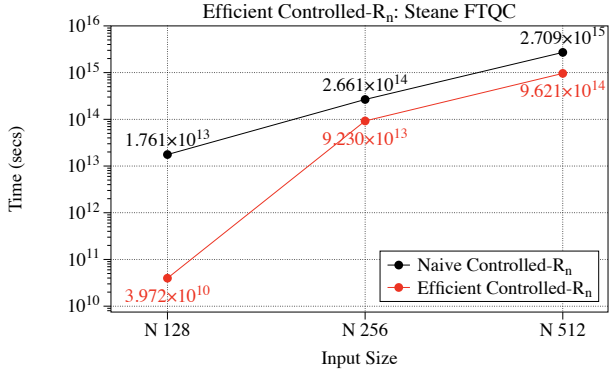, scale=0.22}
}
\subfigure[]{
	\epsfig{file=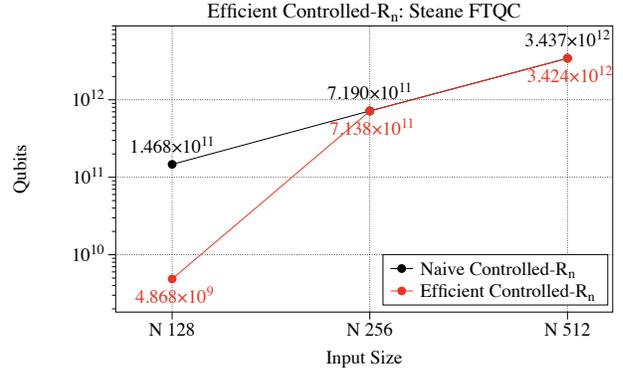, scale=0.22}
}
\caption
{
The performance comparison between a naive compile and the proposed compile under Steane code based quantum computing.
(a) The algorithm execution time. (b) The required qubits.
}
\label{fig:compile_improvement_steane}
\end{figure*}

FIG.~\ref{fig:compile_improvement_steane} shows the performance improvement by the efficient compile in the Steane code quantum computing.
At the input size $N=128$, the improved decomposition lowers the quantum computing time as much as over 400-fold and the qubits as much as 30-fold.
The degree of the performance improvement depends on the input size.
As shown in the figure, at the input size where the required concatenation level lowers by applying the proposed decomposition, the performance improvement is remarkable.

\begin{figure*}[t]
\centering
\subfigure[]{
	\epsfig{file=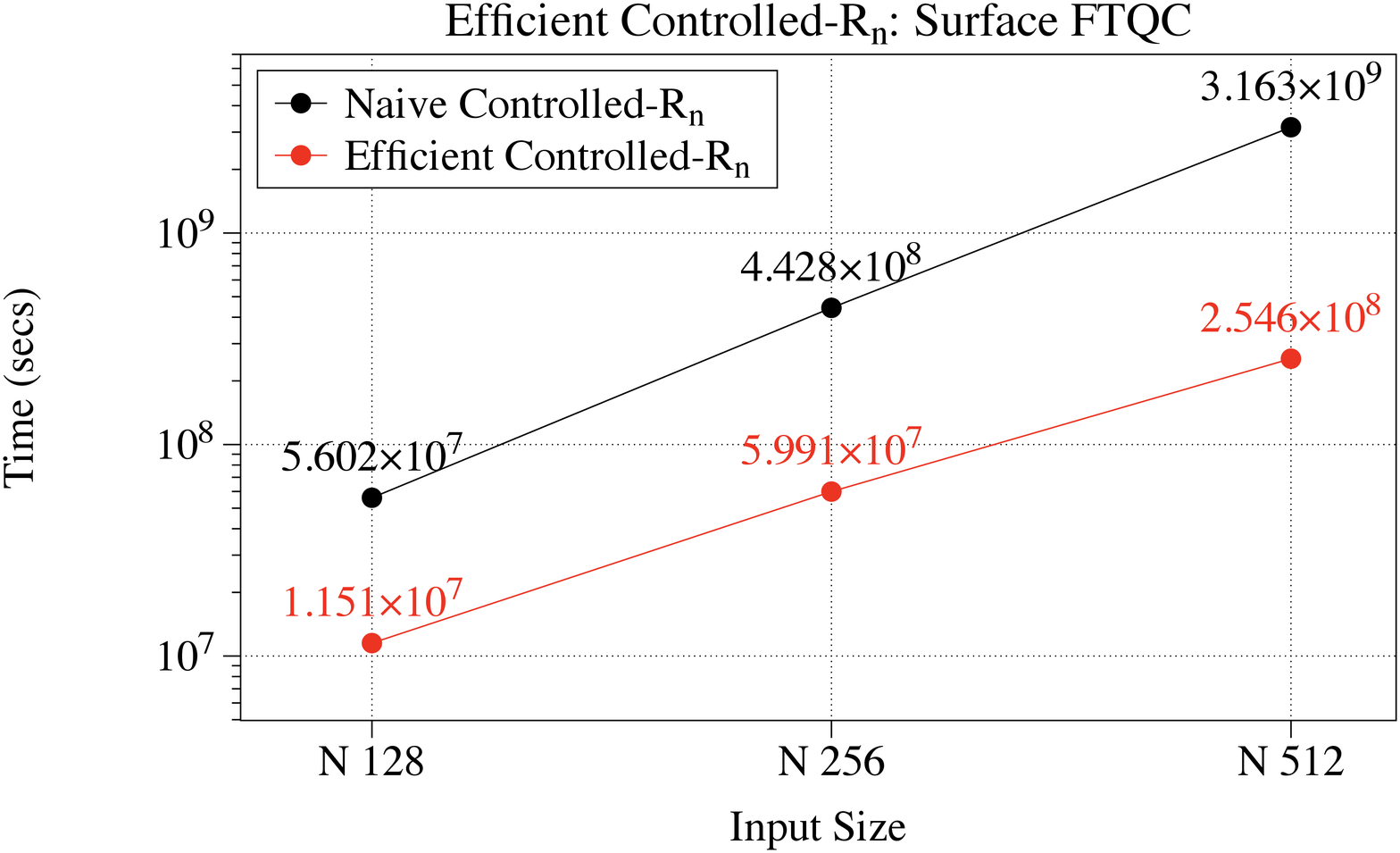, scale=0.22}
}
\subfigure[]{
	\epsfig{file=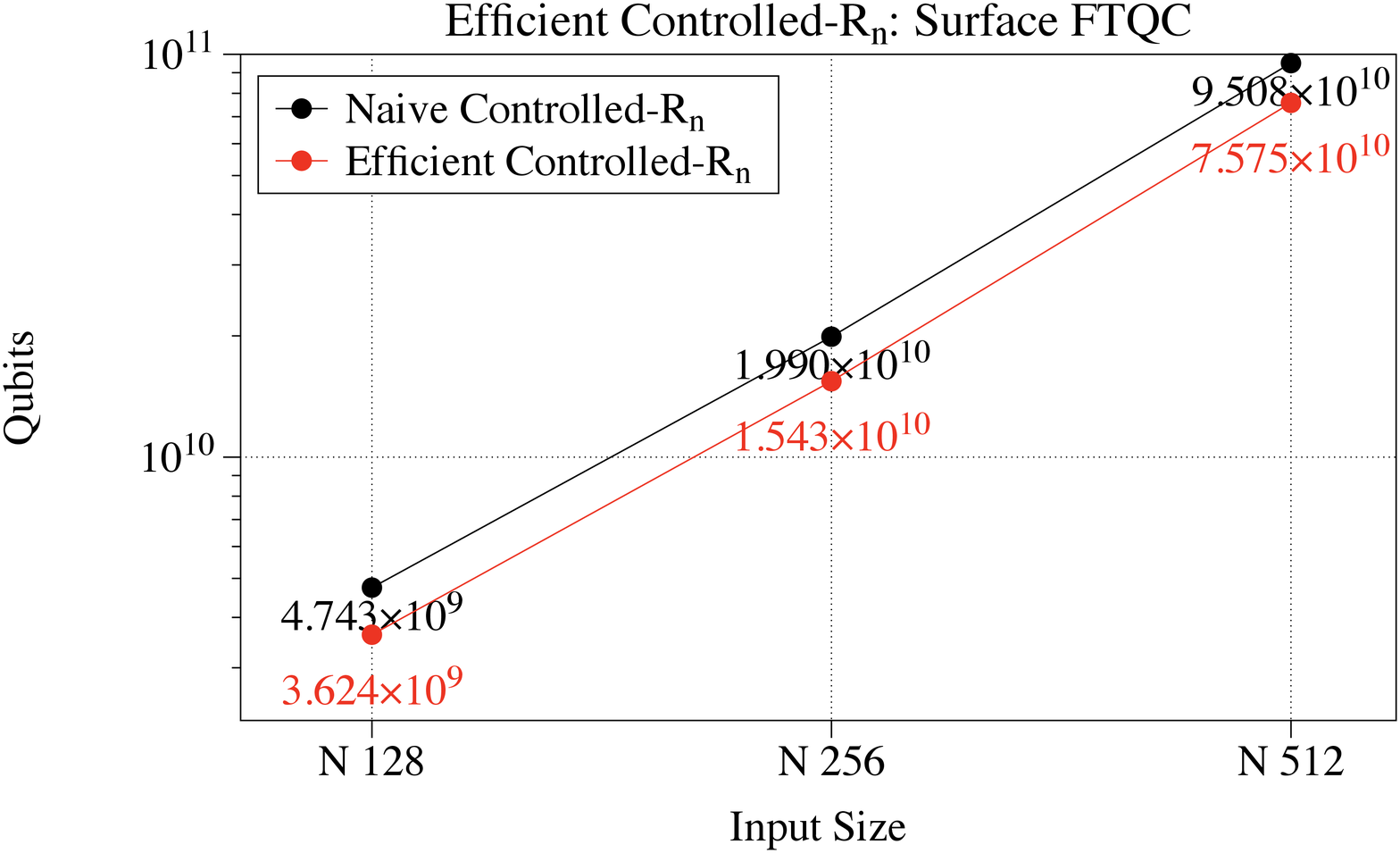, scale=0.22}
}
\caption
{
The performance comparison between a naive compile and the proposed compile under surface code based quantum computing.
(a) The algorithm execution time and (b) The required qubits.
}
\label{fig:compile_improvement_surface}
\end{figure*}

FIG.~\ref{fig:compile_improvement_surface} shows the performance improvement by the efficient compile in the surface code quantum computing.
Unlike the Steane code quantum computing, the performance improvement in the quantum computing time increases gradually as the input size increases.
This is because there always exists the difference in the code distance.
By the improved decomposition, the required code distance lowers to 22, 24, 27 from 25, 27, 30 respectively.

\subsection{Accurate Quantum Gates}\label{sec:accurate_gates}
Previously, we basically assumed that physical error rate is $10^{-9}$ for Steane code quantum computing and $10^{-3}$ for surface code quantum computing.
In this section, we show what happens in a quantum computing if we have more accurate quantum device.
For that we show the performance evaluations based on the physical error rates $10^{-9}\sim 10^{-15}$ for Steane code quantum computing and based on the physical error rates over $10^{-3}\sim 10^{-9}$.

\begin{figure*}[t]
\centering
\subfigure[]{
	\epsfig{file=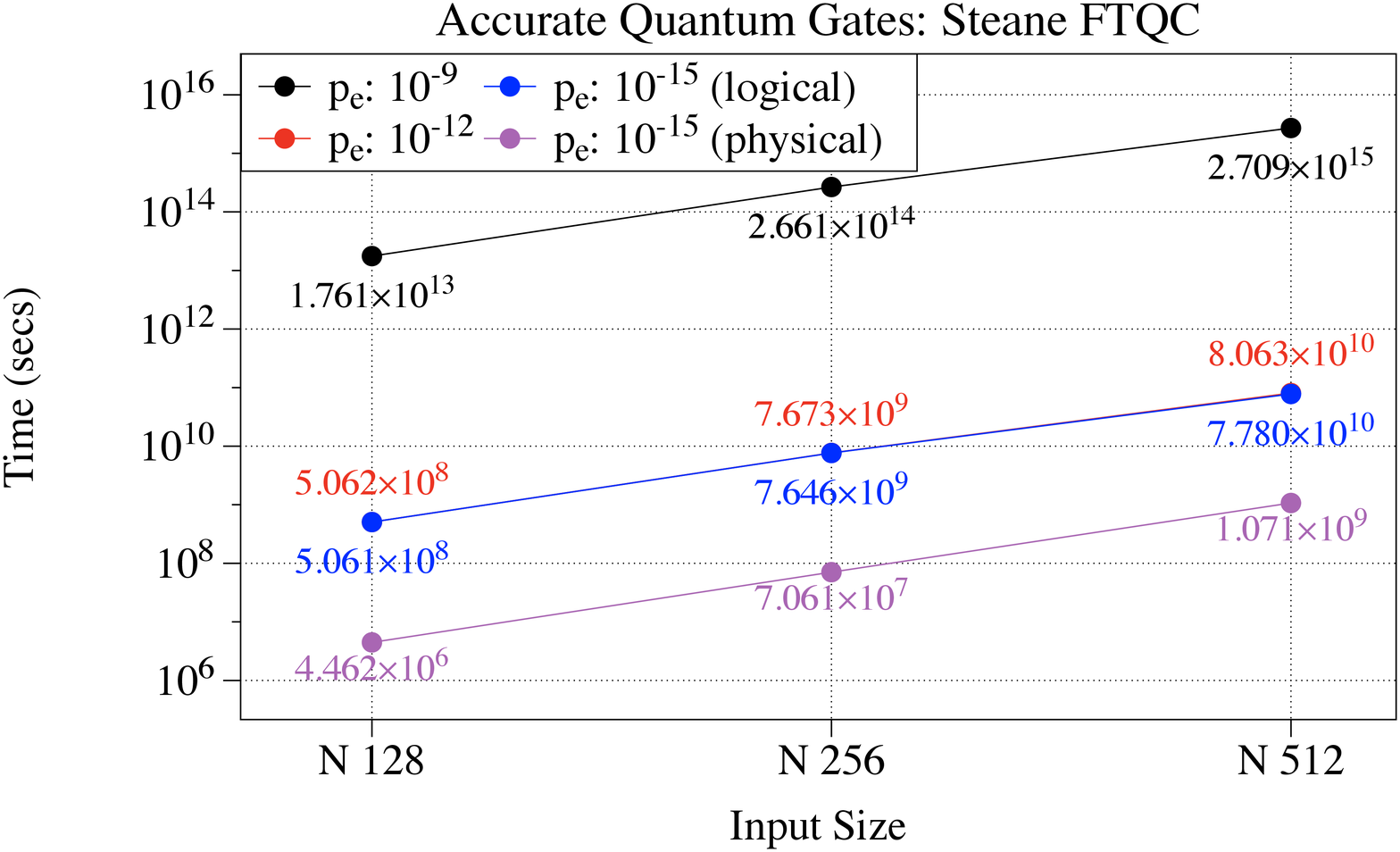, scale=0.22}
}
\subfigure[]{
	\epsfig{file=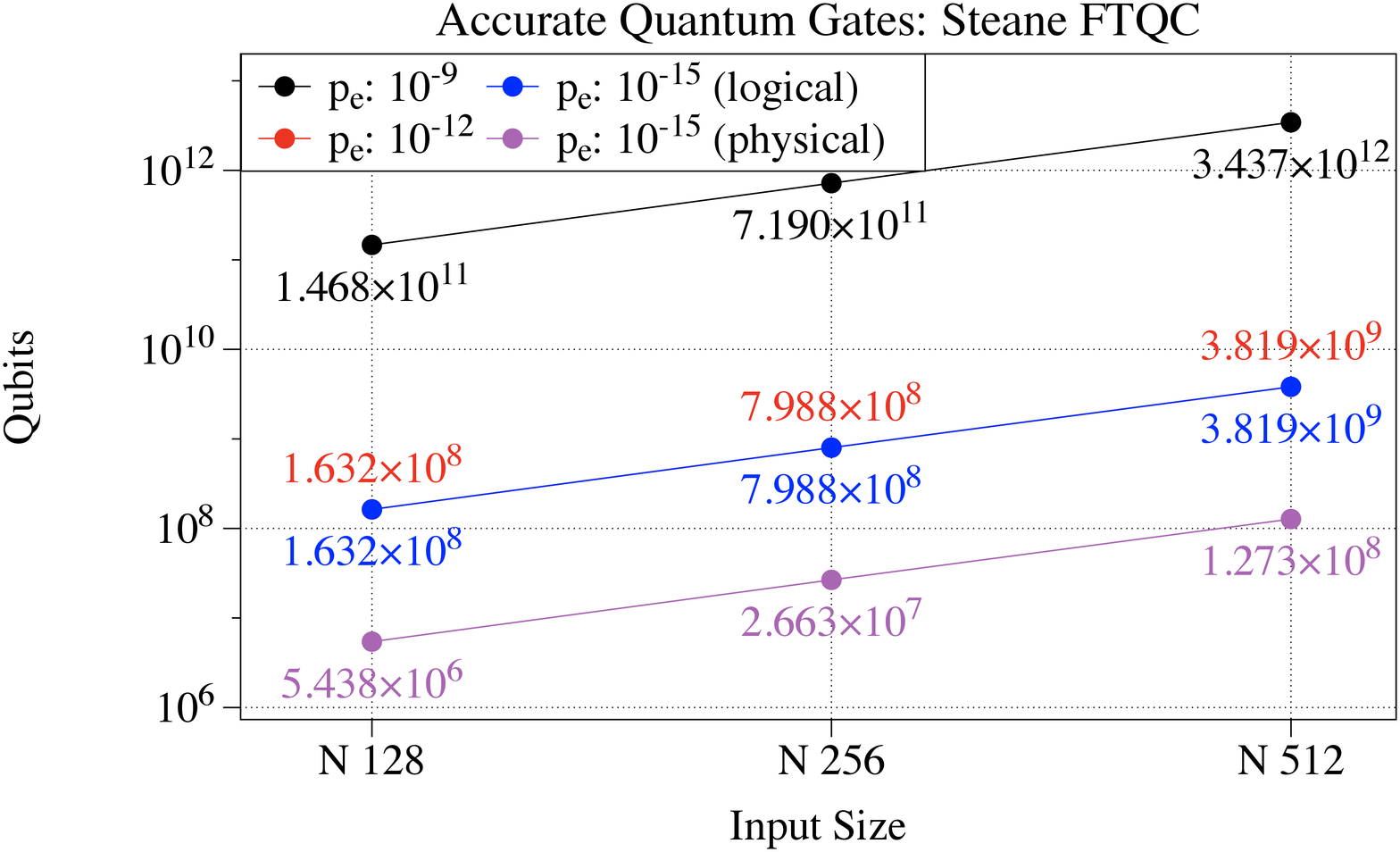, scale=0.22}
}
\caption
{
The performance comparison over physical error rates $10^{-9}\sim 10^{-15}$ under Steane code based quantum computing. 
(a) The algorithm execution time and (b) The required qubits.
At the error rate $10^{-15}$, as shown in this figure, a fault-tolerant quantum computing is not required.
}
\label{fig:robust_gate_steane}
\end{figure*}

FIG.~\ref{fig:robust_gate_steane} shows the Steane code quantum computing performance over physical error rates $10^{-9}\sim 10^{-15}$.
We also compare a physical quantum computing to those fault-tolerant quantum computings at the physical error rate $10^{-15}$.
The performance improvement by lowering the error rate from $10^{-9}$ to $10^{-12}$ is highly nontrivial because the required concatenation level is reduced from 2 and 3 to 1 in both case respectively.
But, lowering the error rate more does not lead to the better fault-tolerant quantum computing performance.
In other words, the fault-tolerant quantum computing in the physical error rate $10^{-15}$ does not show any advantage against the quantum computing in the physical error rate $10^{-12}$.
This is because as the physical error rate lowers the fault-tolerant quantum computings with the same concatenation level achieves very high fidelity ($>0.9$).
If both quantum computings are performed with the same concatenation level, both have the same single round quantum computing time.
In that case, if there is no big difference between fidelities, the average quantum computing $T_{avg}$ will be very similar.

In the same reason, in the physical error rate $10^{-15}$, physical quantum computing shows the better performance than a fault-tolerant quantum computing because the physical quantum computing already achieves high fidelity.
Empirically, $T_{one}$ of the physical quantum computing in the error rate $10^{-15}$ is $6.89 \times 10^{8}$  with the fidelity $0.6433$.
On the other hand, $T_{one}$ of the fault-tolerant quantum computing is $7.78 \times 10^{10}$ with the fidelity $0.9999$.
Obviously, physical quantum computing shows the better performance in terms of $T_{avg}$, $\frac{6.89\times 10^{8}}{0.6433} > \frac{7.78\times 10^{10}}{0.9999}$.


\begin{figure}[t]
\centering
\subfigure[]{
	\epsfig{file=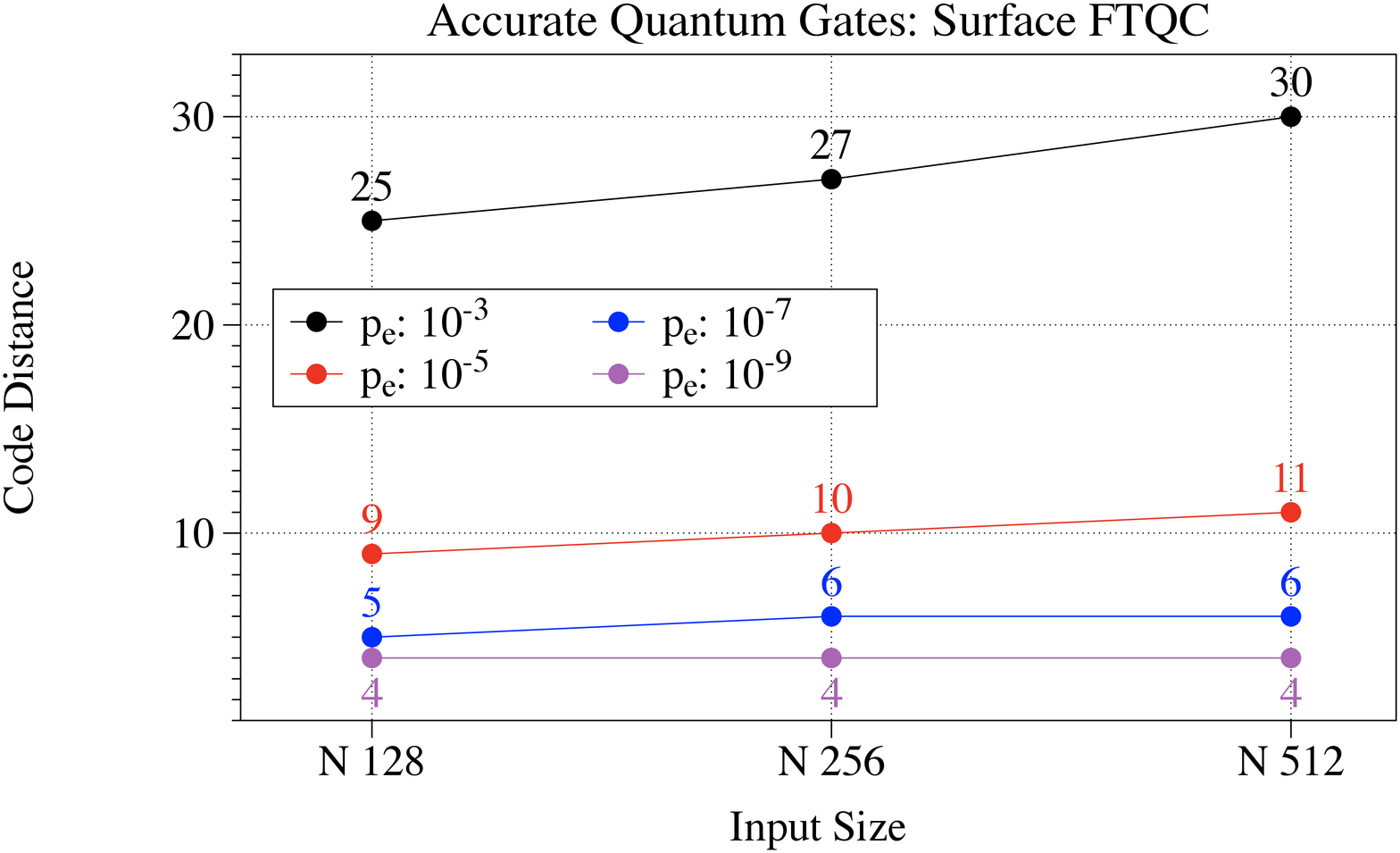, scale=0.22}
}
\subfigure[]{
	\epsfig{file=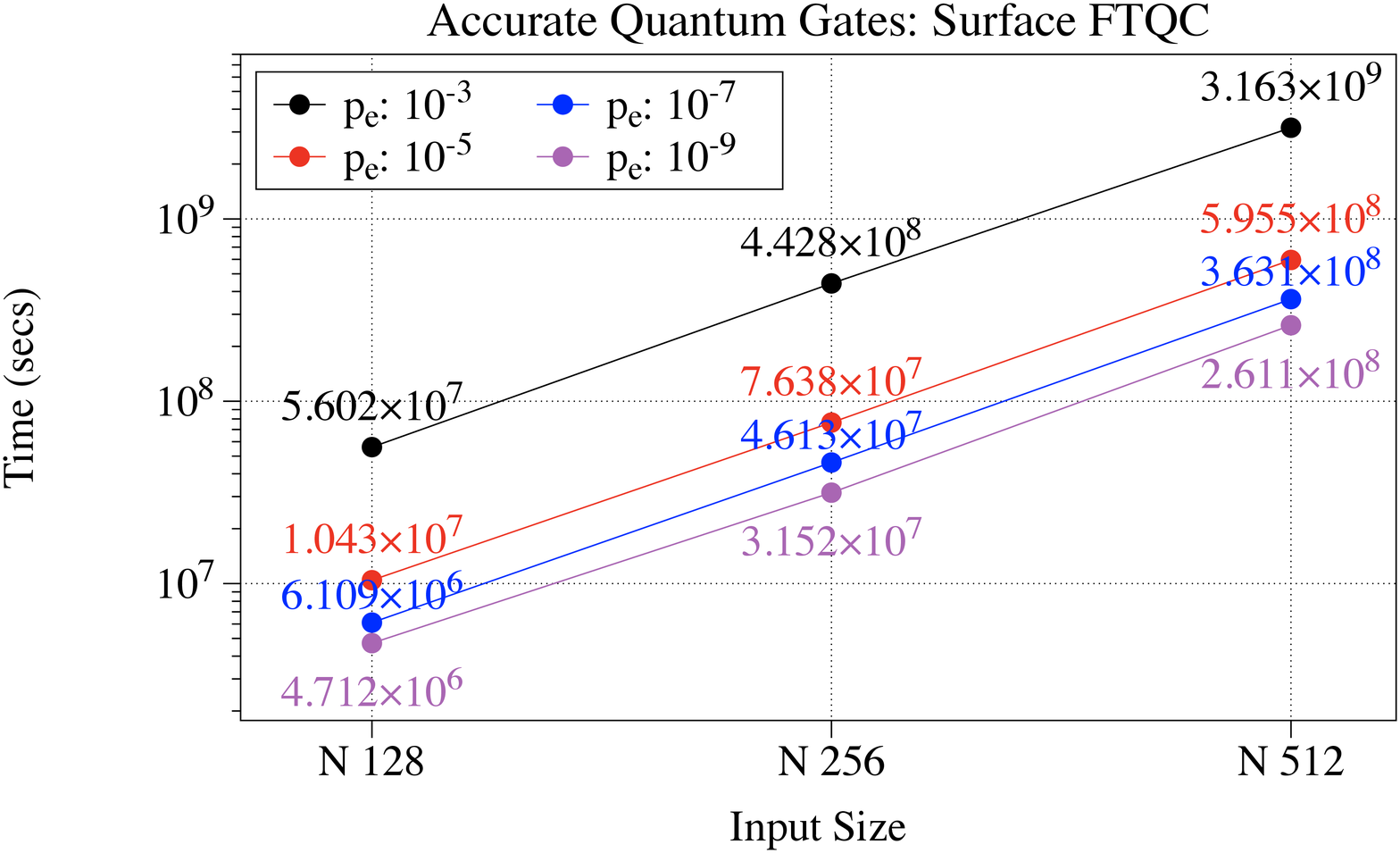, scale=0.22}
}
\subfigure[]{
	\epsfig{file=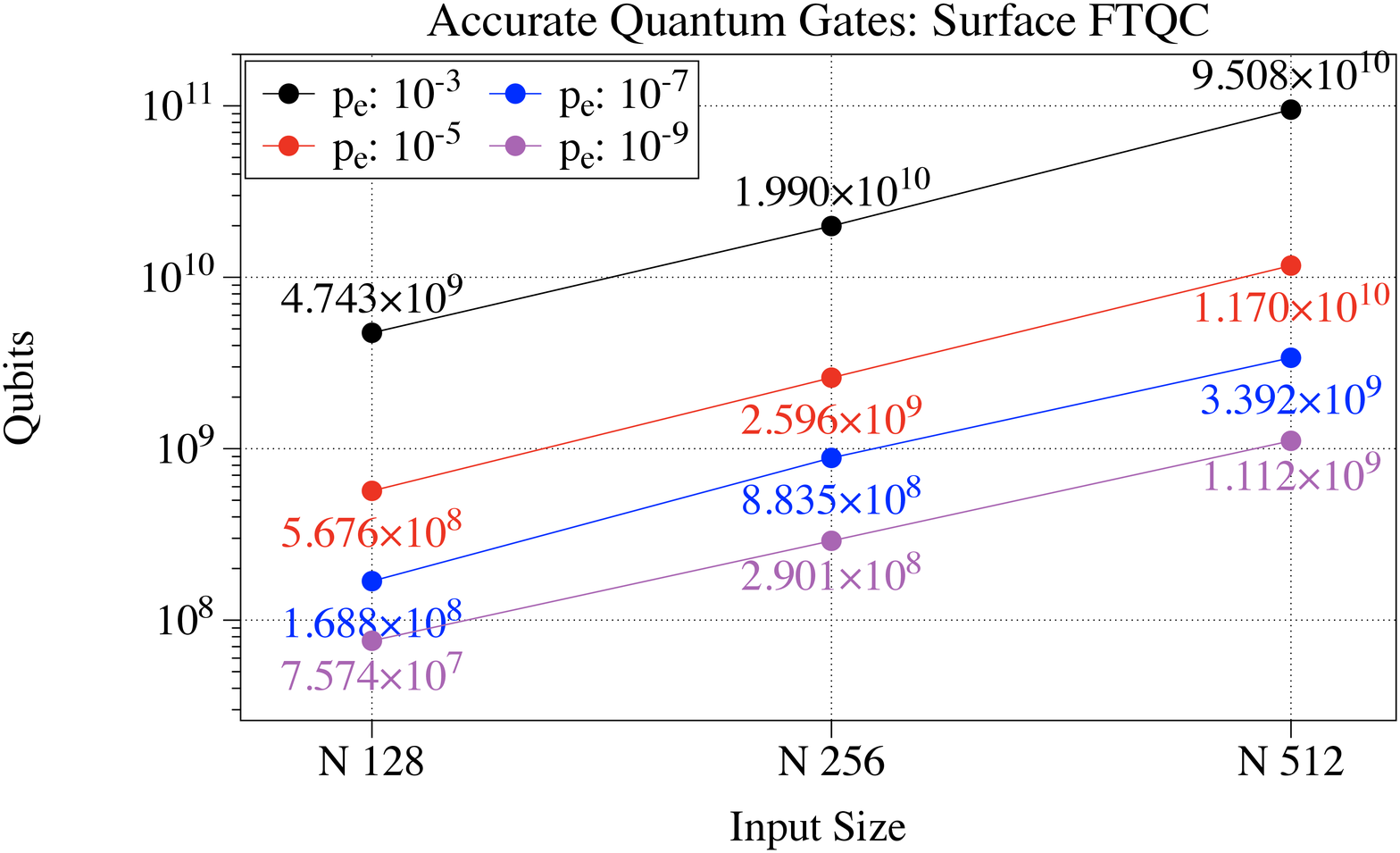, scale=0.22}
}
\caption
{
The performance comparison over physical error rates $10^{-3}\sim 10^{-9}$ under surface code based quantum computing.
(a) The code distance, (b) The algorithm execution time and (c) The required qubits. 
}
\label{fig:robust_gate_surface}
\end{figure}

FIG.~\ref{fig:robust_gate_surface} shows the performance improvement in the surface code quantum computing over physical error rates $10^{-3}\sim 10^{-9}$.
In the figure, we also compare the required code distance.
As shown in the figure, as the physical error rates lowers, the required code distance decreases and therefore the performance increases.
But, since the code distance is already too low, 4 or 5, there is not enough room for the performance improvement as the gate is improved more.

\subsection{Degree of Fault Tolerance}\label{sec:degree_fault_tolerance}
\begin{figure}[t]
\centering
\epsfig{file=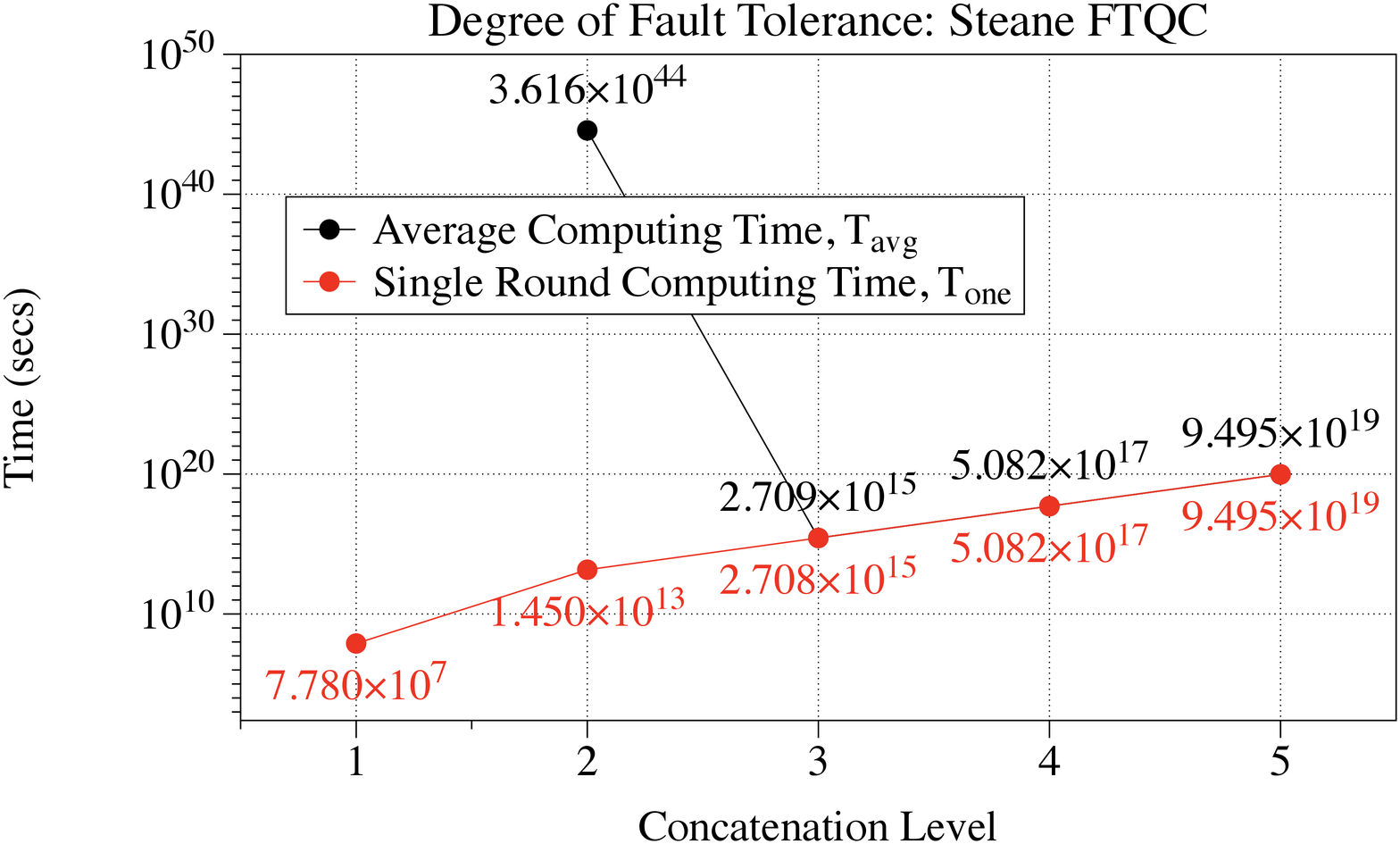, scale=0.23}
\caption{
	The quantum computing time of Shor algorithm with input size $N=512$.
	We have evaluated the quantum computing performance $T_{one}$ and $T_{avg}$ according to the concatenation levels $1\sim 5$.
	After the concatenation level 3, the fidelity of a quantum computing is almost 1 and therefore the average computing time closely approaches to the single round computing time.
	When, the concatenation level is 1, the fidelity of a quantum computing is almost vanishing and therefore the average computing goes to almost infinity.
}
\label{fig:degree_ft_steane_code}
\end{figure}

Accuracy threshold theorem~\cite{Aharonov:2008jn,Knill:1996tm} says that if we have quantum device of physical error rates below a threshold, it is possible to achieve an arbitrary long quantum computation is possible.
By applying a recursive concatenated coding~\cite{Knill:1996ty}, we can lower the effective error rate to where a reliable quantum computing is possible.
As we increase the concatenation level, the fidelity of a quantum computing also definitely improves.
But, the duration of a quantum computing is also increased by raising a concatenation level.
Therefore, the higher concatenation level does not always make the better quantum computing possible.
FIG.~\ref{fig:degree_ft_steane_code} shows that there exists a recommendation for the concatenation level in the Steane code quantum computing, in particular for a quantum computing time.
The number of qubits unconditionally becomes larger as the concatenation level increases.

In case of a surface code quantum computing, the performance completely depends on the code distance.
The code distance is determined to satisfy the objective fidelity of a quantum computing, but in most case the accuracy of the quantum computing by the chosen code distance exceeds than the target fidelity.
In this regards, on considering the averaged quantum computing time $T_{avg}$, the chosen code distance may not bring the best quantum computing performance as shown in the Steane code case.

\begin{figure}[t]
\centering
\epsfig{file=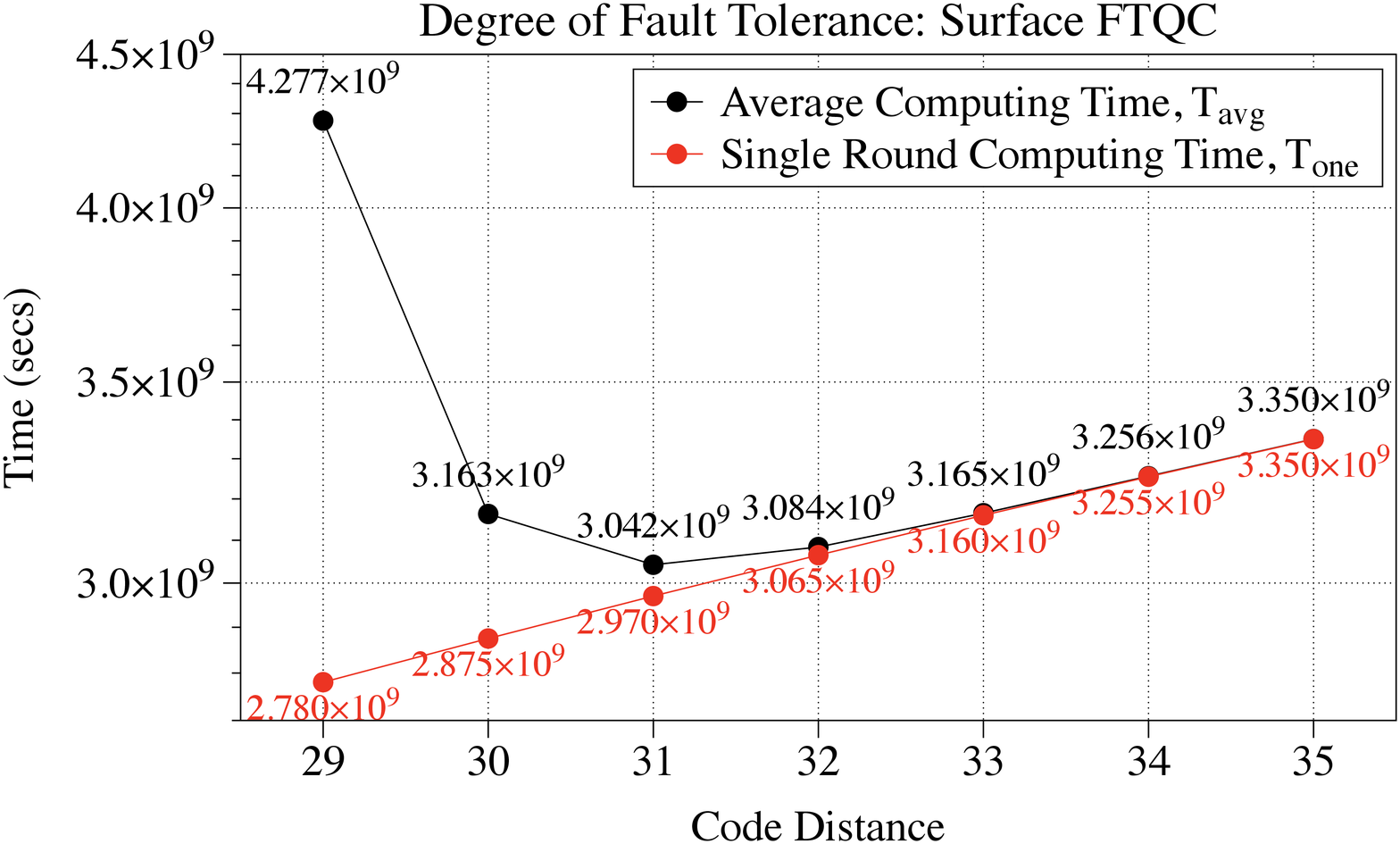, scale=0.23}
\caption{
	The quantum computing time of Shor algorithm with input size $N=512$.
	We have evaluated the quantum computing performance $T_{one}$ and $T_{avg}$ by varying the code distance from 29 to 35.
	The calculated code distance for the objective fidelity is 30. 
	As shown in the figure, the code distance 31 introduces the best quantum computing performance.
	By taking the code distance 31, we can reduce the quantum computing time as much as 1400 days than the case of the distance 30 at the cost of qubits.
}
\label{fig:degree_ft_surface_code}
\end{figure}

FIG.~\ref{fig:degree_ft_surface_code} shows that the surface code quantum computing has the best performance with a code distance $31$, but the code distance determined by the equation is $30$.
Even though the code distance determined from the target fidelity 0.7 is 30, the goal of a quantum computing is to find an exact answer, not a probable answer.
By applying the code distance 31, we can reduce the quantum computing time as much as 1400 days than the case of the code distance 30 at the cost of qubits.

\section{Discussion}\label{sec:discussion}

We have proposed an integrated method for analyzing the performance and the resource of a quantum computing.
In particular, by considering practically running a quantum algorithm on a quantum computer hardware of a specific system architecture, we have obtained the most realistic performance and resource where the effects by all of fully decomposed algorithm, fault-tolerant scheme and system architecture are involved.
For that, we have proposed and developed a quantum computing framework composed of three functional layers where each layer plays a definite role of a quantum computing.
By exploiting the framework, we can configure a quantum computing model by selecting specific protocol and/or properties.
By doing so, we can analyze not only the performance and resource of a quantum computing, but also the impact of specific components on the entire quantum computing.
For example, we have discussed optimal concatenation level and code distance of fault-tolerant quantum computing.
We believe such discussion was possible due to the proposed framework.

The analysis results completely depend on the protocols and properties of physical device we employed.
As shown in the figures, the quantity of the required qubits is too enormous and the execution time is too long.
The feasibility of a quantum computing seems too bad from our analysis results.
However, we need to emphasize that the very those figures are not so important now.
Readers need to see the tendency of the analysis results as the input size of a quantum algorithm increases.
As quantum computing components are being improved, the analysis results will be better than the shown in this work, but the tendency may be stayed.

As mentioned above, the objective of the present work is to provide the most realistic performance and resource of a quantum computing.
On the other hand, we believe the proposed software framework can play a significant role in practically running a quantum computing with a real quantum computing hardware later if some components are added (see FIG.~\ref{fig:data_flow_components}).
For example, components to control a real quantum device have to be added in the building block layer.
The system layer also requires functions that execute a quantum algorithm and a quantum error correction efficiently.
Besides, the existing components have to be improved as much as possible.

\begin{acknowledgements}
This work was supported by Electronics and Telecommunications Research Institute (ETRI) grant funded by the Korean government [18ZH1400, Research and Development of Quantum Computing Platform and its Cost-Effectiveness Improvement].
\end{acknowledgements}


\end{document}